\documentclass[aps,prx,showpacs,superscriptaddress,twocolumn]{revtex4-1} 
\usepackage{graphicx}
\usepackage{bm}
\usepackage{amssymb} 
\usepackage{amsmath} 
\usepackage{amsthm}

\usepackage{hyperref}
\usepackage[utf8]{inputenc}
\hypersetup{
     colorlinks=true,      
    linkcolor=blue,        
    citecolor=blue,
    filecolor=magenta, 
    urlcolor=black          
}

\newcommand{\trc}[1]{\tr\left\{#1\right\}}
\newcommand{\onefrac}[1]{\frac{1}{#1}}
\newcommand{\expi}[1]{\mathbb{E}_{U}\left[#1\right]}
\newcommand{\expubi}[1]{\mathbb{E}_{U,\, \text{multi}}\left[#1\right]}

\newcommand{\exppsiu}[1]{\mathbb{E}_{\psi, \, U}\left[#1\right]}
\newcommand{\exppsi}[1]{\mathbb{E}_{\psi}\left[#1\right]}
\newcommand{\expui}[1]{\mathbb{E}_{u_i}\left[#1\right]}

\newcommand{\expbi}[1]{\mathbb{E}_\text{multi}\left[#1\right]}
\newcommand{\Var}[1]{\text{Var}\left[#1\right]}
\renewcommand{\vec}[1]{\bm{#1}}
\newlength{\leftstackrelawd}
\newlength{\leftstackrelbwd}
\def\leftstackrel#1#2{\settowidth{\leftstackrelawd}%
	{${{}^{#1}}$}\settowidth{\leftstackrelbwd}{$#2$}%
	\addtolength{\leftstackrelawd}{-\leftstackrelbwd}%
	\leavevmode\ifthenelse{\lengthtest{\leftstackrelawd>0pt}}%
	{\kern-.5\leftstackrelawd}{}\mathrel{\mathop{#2}\limits^{#1}}}

\providecommand{\ket}[1]{\lvert #1 \rangle}

\providecommand{\ketbra}[2]{\lvert  #1\rangle \langle #2 \rvert}
\newcommand{\expec}[1]{\langle #1\rangle}
\newcommand{\tr}{\mathrm{tr}\,} 

\usepackage{mathbbol}

\begin{document}

\title{Quantifying multiparticle entanglement with randomized measurements}

\author{Sophia Ohnemus}
\affiliation{Physikalisches Institut, Albert-Ludwigs-Universit\"at Freiburg, Hermann-Herder-Str. 3,
79104 Freiburg, Germany}

\author{Heinz-Peter Breuer}
\affiliation{Physikalisches Institut, Albert-Ludwigs-Universit\"at Freiburg, Hermann-Herder-Str. 3,
79104 Freiburg, Germany}
\affiliation{EUCOR Centre for Quantum Science and Quantum Computing, Albert-Ludwigs-Universit\"at Freiburg, Hermann-Herder-Str.~3, 79104~Freiburg, Germany}

\author{Andreas Ketterer}
\email{andreas.ketterer@iaf.fraunhofer.de}
\affiliation{Physikalisches Institut, Albert-Ludwigs-Universit\"at Freiburg, Hermann-Herder-Str. 3,
79104 Freiburg, Germany}
\affiliation{EUCOR Centre for Quantum Science and Quantum Computing, Albert-Ludwigs-Universit\"at Freiburg, Hermann-Herder-Str.~3, 79104~Freiburg, Germany}
\affiliation{Fraunhofer Institute for Applied Solid State Physics IAF, Tullastr.~72, 79108~Freiburg, Germany}

\begin{abstract}
Randomized measurements constitute a simple measurement primitive that exploits the information encoded in the outcome statistics of samples of local quantum measurements defined through randomly selected bases. In this work we exploit the potential of randomized measurements in order to probe the amount of entanglement contained in multiparticle quantum systems as quantified by the multiparticle concurrence. We further present a detailed statistical analysis of the underlying measurement resources required for a confident estimation of the introduced quantifiers using analytical tools from the theory of random matrices. The introduced framework is demonstrated by a series of numerical experiments analyzing the concurrence of typical multiparticle entangled states as well as of ensembles of output states produced by random quantum circuits. Finally, we examine the multiparticle entanglement of mixed states produced by noisy quantum circuits consisting of  single- and two-qubit gates with non-vanishing depolarization errors, thus showing that our framework is directly applicable in the noisy intermediate-scale regime.
\end{abstract}

\maketitle 
\section{Introduction} 
The multiparticle entanglement content of composite quantum states of many, possibly interacting particles plays a central role for the development of novel quantum technologies, ranging from quantum  communication protocols to quantum computing architectures~\cite{NielsenChuang,Buhrman:2010uq,QuSup,SupercondQubitsRev}. For instance, multiparticle entanglement has been shown to enhance the performance of anonymous conference key agreement \cite{Graselli2021}, act as a resource in quantum metrology~\cite{Gessner2021}, and it is believed to be a crucial ingredient for quantum computation algrorithms outperforming analogous classical counterparts~\cite{VanDenNest2013}. However, determining a state's content of multiparticle entanglement becomes increasingly difficult with growing particle number due to the large dimension and immense complexity of the underlying multiparticle Hilbert space~\cite{H4EntReview,OtfriedReview}. 

Methods for the characterization of multiparticle properties vary strongly in terms of the required measurement resources as well as their assumptions upon the states under considerations~\cite{ReviewCertification}. While tomographic tools assume very little about the considered quantum states they become impractical already for rather small system sizes~\cite{TomographyReview}. In contrast, witness operators allow for an efficient certification of multiparticle properties, such as structures of entanglement or the states' fidelities, but their successful implementation relies heavily on the knowledge of the investigated quantum states~\cite{OtfriedReview,SelfTestingRev,FidelityWitnesses}. Other compromises between these two extreme strategies allow to lower the required measurement resources by either invoking specific prior information about the sparsity of the involved density operators~\cite{CompressedSensing1,CompressedSensing2} or by accepting limited precision requirements of the targeted observables~\cite{LeandroRegenerativeModels}. 

\begin{figure}[t!]
\begin{center}
\includegraphics[width=0.47\textwidth]{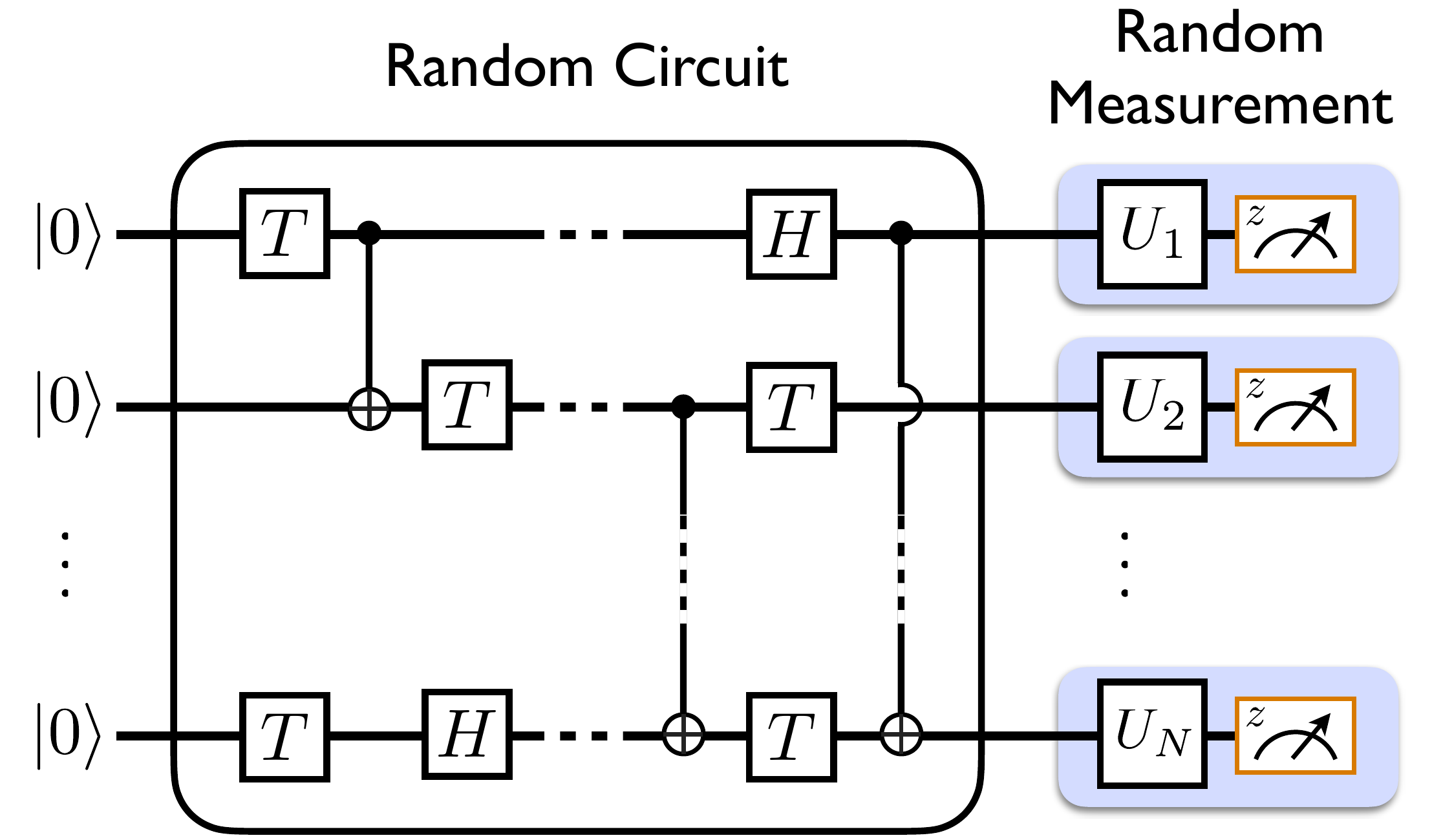}%
\end{center}
\caption{Random quantum circuit consisting of $N$ qubits. The qubits are initialized in the ground state $\ket{0}^{\otimes N}$ and subsequently manipulated with gates drawn form the  universal gate set $\mathcal I_\text{uni}=\{H,T,C_X\}$ which are applied to randomly selected qubits. The resulting output state of the computation is analysed using a randomized measurement protocol consisting of randomly drawn local unitary transformations $U_i$ complemented with a measurement in the computational basis of $N$ qubits.
}
\label{fig_1}
\end{figure}

A promising approach in this regard is based on so-called randomized measurements where the underlying quantum state is readout in randomly selected local measurement bases and system properties are inferred via appropriate statistical averages~\cite{RudolphRandMeasBell,FlammiaFidelityStat,BrunnerRandMeasBell,EnkPRL2012,tran1,tran2,Mattia2016,Giordani2018,MeMoments1,MeMoments2,MichaelBachelor,ZollerFirst,vermerschPRA,ZollerScience,ElbenPRA,DakicRandomMeas,MeineckeExperimentRandom,ElbenPRL,ElbenPRLmixedstate,SatoyaMoments,MeMoments3,RandTriads,KnipsPerspective} (see Fig.~\ref{fig_1}). In this way it has been shown to be possible to extract a number of relevant properties of the underlying many-body states such as structures of multiparticle entanglement~\cite{MeMoments1,MeMoments2,SatoyaMoments,MeMoments3}, subsystem purities~\cite{ZollerScience,ElbenPRA}, fidelities with respect given target states or other quantum devices \cite{FlammiaFidelityStat,ElbenPRL}, or interference signatures of indistinguishable particles~\cite{Mattia2016,Giordani2018}. Furthermore, the underlying measurement resources for a statistically significant verification of the aforementioned properties have been investigated, promising advantages particularly in the noisy intermediate-scale regime~\cite{Shadows2,ElbenPRLmixedstate,MeMoments3}.
 
In this work we use locally randomized measurements in order to directly extract information about the amount of entanglement in terms of the multiparticle concurrence~\cite{Carvalho2004,Mintert2005a,Mintert2005a,Mintert2005b,AolitaConc1,AolitaConc2}, a quantifier for multiparticle entanglement. In particular, we derive an exact formula for the multiparticle concurrence of pure states, as well as for an appropriate lower bound in the case of mixed states, from second moments of the outcomes of random measurements. Furthermore, we analyze in detail the measurement resources required for a statistically confident estimation of the involved quantities by analysing the respective variances. We apply the developed toolbox to evaluate the multiparticle entanglement of typical multiparticle states as well as of ensembles of states produced by different classes of random quantum circuits (see Fig.~\ref{fig_1}). Finally, we investigate the multiparticle entanglement of mixed states produced by noisy quantum circuits consisting of gates prone with non-vanishing depolarization errors.

The paper is structured as follows: In Sec.~\ref{sec:RandMeas} we introduce the paradigm of randomized measurements and show how it enables a measurement of the concurrence of pure multiparticle quantum states as well as of a suitable lower bound in the case of mixed states. Further on, in Sec.~\ref{sec:EstStatErr} we discuss the statistical estimation of the involved quantities based on finite samples of randomized measurements and, in particular, analyse the involved statistical error through evaluation of the variances of the respective estimators. In Sec.~\ref{sec:ApplEntStates} and \ref{sec:RandCirc}, respectively, we demonstrate the introduced protocols through numerical simulations of typical examples of multiparticle entangled states as well as of ensembles of states produced by random quantum circuits. Finally, we conclude our work in Sec.~\ref{sec:Conclusion} and give a brief outlook.
\section{Probing multiparticle entanglement with randomized measurements}\label{sec:RandMeas}
\subsection{Randomized measurements and moments of random correlations}\label{sec:RandMoments}

To start with we introduce briefly the framework of randomized measurements as a diagnostic tool for the characterization of multiparticle quantum systems. For reasons of generality we consider in this section a collection of $N$ $d$-dimensional quantum systems (qudits) each described by a local Hilbert space $\mathcal H=\mathbb C^d$. The quantum state of the total multiparticle particle system is then described by a density operator $\varrho$ acting on the $N$-particle Hilbert space $\mathcal H^{\otimes N}$. 

A random measurement of the $N$-particle state $\varrho$ is then described through a set of randomly drawn local bases $\{ \mathcal B_n\}_{n=1,\ldots,N}$, each defined as $\mathcal B_n=\{ U_n\ket{s_n}\}_{s=0,\ldots,(d-1)}$ 
with a random transformation $U_n$ picked uniformly from the unitary group $\mathcal U(d)$, i.e., according to the Haar measure, and where $\{\ket{s_n}\}_{s=0,\ldots,(d-1)}$ denotes the computational basis of the $n$'th qudit. The outcome of a single measurement run in  such a random basis is then  labelled by a string $\boldsymbol s=(s_1,\ldots,s_N)$ of length $N$ containing values $s_k=0,\ldots,d-1$ and the associated outcome probability reads $P_U(\boldsymbol s)=\text{tr}[\varrho U\ketbra{\boldsymbol s}{\boldsymbol s} U^\dagger]$, with $U=U_1 \otimes \dots \otimes U_N$. 

The idea behind randomized measurement protocols is now to regard appropriate combinations of the population probabilities $P_U(\boldsymbol s)=\text{tr}[\varrho U\ketbra{\boldsymbol s}{\boldsymbol s} U^\dagger]$ in such a way that, upon averaging them uniformly over the local unitary group $\mathcal U(d)^{\otimes N}$, they provide insights about the properties of the quantum state $\varrho$. For instance, it has been shown in Refs.~\cite{ZollerScience,ElbenPRA} that the state's purity can be obtained as follows:
\begin{align}
\tr[\varrho^2]=d^{N} \sum_{\mathbf{s},\mathbf{s}'} (-d)^{-D(\mathbf{s}, \mathbf{s}')} \overline{P_U(\mathbf{s})P_U(\mathbf{s}')},
\label{eq:PurityRandMeas}
\end{align}
where $D(\boldsymbol s,\boldsymbol s')=\# \{i\in \{1,\ldots,N\}|s_i\neq s_j\}$ denotes the Hamming distance between two computational basis states $\ket{\boldsymbol s}=\ket{s_1,\ldots,s_N}$ and $\ket{\boldsymbol s'}=\ket{s_1',\ldots,s_N'}$.

Analogously, we can associate a random measurement of a single qudit with an observable $\mathcal{O}_U=U\mathcal OU^\dagger$, where $U\in \mathcal U(d)$ and  
$\mathcal O$ denotes a traceless observable diagonal in the computational basis with outcomes $\{o_i\}_{i=0,\ldots,(d-1)}$. For instance, for qubits $(d=2)$ a standard choice of the observable $\mathcal O$ is given by the Pauli matrix $\sigma_z$, leading to the random Pauli matrix $\sigma_{\boldsymbol u_i}$, with $[\boldsymbol u_i]_j=\tr{[\sigma_j U_i\sigma_zU_i^\dagger]}$.
One round of such a random measurement of $N$ qudits thus allows one obtain the correlation functions 
\begin{align}
\expec{ \mathcal{O}^{(i_1)}_{U_{i_1}} \cdot \ldots \cdot \mathcal{O}^{(i_k)}_{U_{i_k}}}=\sum_{\boldsymbol s} o_{s_{i_1}} \ldots o_{s_{i_k}} P_U(\boldsymbol s),
\label{eq:CorrelationFcts}
\end{align}
with a subset $A=\{i_1,\ldots,i_k\}\subset \{1,\ldots,N\}$ of the $N$ qudits of cardinality $k$. Note that 
Eq.~(\ref{eq:CorrelationFcts}) amounts to a classical post-processing of the outcome probabilities $P_U(\boldsymbol s)$ which simplifies significantly in the case of binary observables ($o_s=\pm 1$) where one has to consider overall only two cases, namely those where the parity of the outcomes is either even or odd~\cite{MeMoments3}. Repeating the above measurement strategy many times for  randomly selected choices of the $U_i$'s then results in a distribution of values which encodes the correlations properties of the state $\varrho$ and is characterized by the moments 
\begin{align}
\mathcal R_A^{(t)}= \int_{\mathcal U(d)} d\eta(U_{1}) &\ldots \int_{\mathcal U(d)} d\eta(U_{k})  \expec{ \mathcal{O}^{(i_1)}_{U_{1}} \cdot \ldots \cdot \mathcal{O}^{(i_k)}_{U_{k}}}^t,
\label{eq:RandomMomentsQudits}
\end{align}
where $t$ is a positive integer and $\eta$ the Haar measure on the unitary group $\mathcal U(d)$. 

The moments~(\ref{eq:RandomMomentsQudits}) have been previously shown to be good candidates for the characterization of  multiparticle correlations~\cite{tran1,MeMoments1,MeMoments2,SatoyaMoments,MeMoments3}. In particular, it was shown that the combination of moments of different order leads to an improved sensitivity in the sense that a larger class of states can be detected~\cite{MeMoments1,MeMoments2,SatoyaMoments}. Furthermore, it is often useful to combine moments evaluated on different subsets $A$ of qudits in order to obtain more information about the underlying state (see Refs.~\cite{ElbenPRA,NikolaiSectorLengths,MeineckeExperimentRandom,SatoyaMoments}). For instance, the purity formula~(\ref{eq:PurityRandMeas}) can be expressed as a sum over second order reduced moments of all subsets of qudits, yielding
\begin{align}
    \mathrm{tr}(\varrho^2)= \frac{1}{d^{N}} \sum_{A\subset \{1,\ldots,N\}}(d^2-1)^{|A|} \mathcal R_A^{(2)}(\varrho).
 \label{eq:PuirtyMoments}
\end{align}
where $\mathcal R_{\emptyset}^{(2)}\equiv 1$, with $\emptyset$ denoting the empty set.

In this spirit we will proceed in the following and show that the multiparticle entanglement content, as quantified by multiparticle concurrence, can be assessed via such a randomized measurement protocol.

 \subsection{Multiparticle concurrence from randomized measurements}\label{sec:ConcRandMeas}

The multiparticle concurrence of an $N$ qudit pure state $\ket\psi \in \mathcal H^{\otimes N}$ has been introduced in Refs.~\cite{Carvalho2004,Mintert2005a} and can be expressed as follows
\begin{equation}
C_N(\ket\psi) = 2\ \sqrt{1 - \frac{1}{2^N}\sum_{A \subseteq \{1,\ldots,N\}} \tr{( \varrho_A^2)}},
\label{eq:concurrence}
\end{equation}
where $\varrho_A=\text{tr}_{{A}^c}[\ketbra{\psi}{\psi}]$, with ${A}^c=\{1,\ldots,N\} \setminus A$, denotes the reduced density matrix of the pure state $\ket{\psi}$ with respect to the subsystem associated to the subset $A$ of the $N$ qudits. In App.~\ref{app:A1} we show that $C_N(\ket\psi)$ can be inferred through the following quantity
\begin{equation}
C_N(\ket\psi) = 2\ \sqrt{1 -\frac{d^N(d+1)^N}{2^N} \mathbb E_U\big[{P_{U}^2(\vec{s})}\big]},
\label{eq:ConcPure}
\end{equation}
where $\mathbb E_U[\ldots]$ denotes the average over the ensemble of local unitary transformations $U= U_1 \otimes \dots \otimes U_N$, with $U_i \in \mathcal{U}(d)$, with respect to the local Haar measures on $\mathcal U(d)$. Hence, we have found an expression of the multiparticle concurrence~(\ref{eq:concurrence}) that involves only quantities accessible via randomized measurements and thus avoids the evaluation of the purities of all possible partitions of the $N$ involved subsystems. We note that expression~(\ref{eq:ConcPure}) has been noted previously in Refs.~\cite{MileGu2021RandMeas,MScSophia}. Alternatively, we can express the concurrence in terms of the moments~(\ref{eq:RandomMomentsQudits})  (see App.~\ref{app:A1} for more details), yielding 
\begin{align}
C_N(\ket\psi) = 2\ \sqrt{1-\sum_{A \subset \{1, \ldots, N \}}  \sum_{A' \subset A} \frac{(d^2-1)^{|A'|}}{2^Nd^{|A|}} \mathcal R^{(2)}_{A'}}.
\label{eq:ConcurrenceMoments2}
\end{align}

Further on, generalizations of Eq.~(\ref{eq:ConcPure}) to the case of mixed states usually involve a convex roof construction of the form $C(\varrho)=\inf_{\{p_k,\ket{\phi_k}\}}\sum_k p_k C(\ket{\phi_k})$, where the infimum has to be taken over all possible decompositions $\varrho=\sum_k p_k \ketbra{\phi_k}{\phi_k}$, which is hard to evaluate in practise. This problem can be partially circumvented by considering an appropriate lower bound of the mixed state concurrence $C_N(\varrho)$ as has been derived in Refs.~\cite{Mintert2004,Mintert2005a,Mintert2005b,AolitaConc1,AolitaConc2} leading to the expression  
\begin{equation}
C_N(\varrho) \geq \sqrt{\tr(\varrho\otimes\varrho V_N)},
\label{eq:ConcLowerBound}
\end{equation}
with
\begin{align}
V_N &= 4\left[\mathbf{P}_+ - P_+\otimes\dots\otimes P_+ - (1-2^{1-N})\, \mathbf{P}_-\right],
\end{align}
where $\mathbf{P}_+$ ($\mathbf{P}_-$) denotes the projector onto the (anti-)symmetric subspace of the two-fold tensor copy space $(\mathbb{C}^d)^{\otimes N}\otimes (\mathbb{C}^d)^{\otimes N}$, and similarly $P_+$($P_-$) of the individual qudit subspaces. In App.~\ref{app:A1} we thus show that the lower bound~(\ref{eq:ConcLowerBound}) can be expressed as follows
\begin{align}
C_N(\varrho)^2 &\geq \frac{4}{2^N}-2^{2-N}d^N(d+1)^N \mathbb E_U\left[{P^2_{U}(\vec{s})}\right] \nonumber \\
&+\big(4-\frac{4}{2^{N}}\big)  d^{N} \sum_{\mathbf{s},\mathbf{s}'} (-d)^{-D(\mathbf{s}, \mathbf{s}')} \mathbb E_U\left[{P_U(\mathbf{s})P_U(\mathbf{s}')}\right],
\label{eq:ConcMixed}
\end{align}
with the Hamming distance $D(\boldsymbol s,\boldsymbol s')$. Note that the last term in Eq.~(\ref{eq:ConcMixed}) can be identified as the purity of  $\varrho$ (see Eq.~(\ref{eq:PurityRandMeas})) according to the randomized measurement framework introduced in Refs.~\cite{ZollerFirst,vermerschPRA,ZollerScience,ZollerScience}. Again, we can express Eq.~(\ref{eq:ConcMixed}) equivalently in terms of the moments~(\ref{eq:RandomMomentsQudits}) by combining Eqs.~(\ref{eq:PurityRandMeas}) and (\ref{eq:ConcurrenceMoments2}), leading to
\begin{align}\label{eq:ConcMixedMoments2}
C_N(\varrho)^2&\geq2(1-2^{1-N}) \times \\
\times & \sum_{A \subset \{1, \ldots, N \}} \Big\{\sum_{A' \subset A} \frac{3^{|A'|}}{2^{|A|}} \mathcal R^{(2)}_{A'} +\frac{2\times 3^{|A|}}{2^N}  \mathcal R^{(2)}_A \Big\}. \nonumber
\end{align}
Note that an evaluation of the concurrence or its lower bound through finite samples of randomized measurements using Eqs.~(\ref{eq:ConcPure}) or  (\ref{eq:ConcMixed}), or equivalently using Eqs.~(\ref{eq:ConcurrenceMoments2}) or  (\ref{eq:ConcMixedMoments2}), requires the definition of appropriate unbiased estimators which come with a non-vanishing statistical error (see Sec.~\ref{sec:EstStatErr}).

\subsection{Exact evaluation with quantum designs}\label{sec:Designs}

The above introduced formulas for the evaluation of the multiparticle concurrence based on randomized measurement also provide the starting point for the derivation of exact expressions allowing to determine the concurrence based on a finite number of measurement settings. To do so, we exploit the concept of unitary designs which provide finite sets of unitary matrices that are inequivalent to Haar random ones as long as one is concerned with statistical moments of some finite order. 

Formally, a {unitary $t$-design} is a set of unitary matrices $\{U_k|k=1,\ldots,K^{(t)}\}\subset \mathcal U(d)$, with cardinality $K^{(t)}$~\cite{Dankert}, such that
\begin{equation}
\frac{1}{K^{(t)}} \sum_{k=1}^{K^{(t)}} P_{t',t'}(U_k) =\int_{\mathcal U(d)} P_{t',t'}(U) d\eta(U),
\label{eq:UnitaryDes}
\end{equation}
for all homogeneous polynomials $P_{t',t'}\in \mathrm{Hom}(t',t')$, with $t'\leq t$, and where $\eta(U)$ denotes the normalized Haar measure on $\mathcal U(d)$. We note that $P_{r,s}(U)$ is an element of the set of all homogeneous polynomials $\mathrm{Hom}(r,s)$, with support on the space of unitary matrices $\mathcal U(d)$, that is of degree at most $r$ and, respectively, $s$ in each of the matrix elements of $U$ and their complex conjugates.  
While the existence of unitary designs has been proven \cite{existence}, no universal strategy for their construction in case of an arbitrarily given $t$ is known. However, a number of approximate unitary designs, for which the property (\ref{eq:UnitaryDes}) is accordingly relaxed, have been introduced in the literature~\cite{ApproxDesign1,ApproxDesign2,ApproxDesign3}. In the remainder of this manuscript we will restrict ourselves to the particular case of the Clifford group $\mathcal C(d)\subset \mathcal U(d)$ which has been shown to constitute a unitary $3$-design~\cite{Cliff3Design,CliffNo4Design}. The Clifford group consists in general of all unitary matrices that map the multi-qudit Pauli group onto itself. In the case of a single qubit this amounts to $|\mathcal C(2)|=24$ elements which can be generated from the Hadamard gate $H$ and the phase gate $ S=e^{i \frac{\pi}{4} \sigma_z}$. 

Now, noting that the second power of the correlation function (\ref{eq:CorrelationFcts}) is a polynomial of degree two in the entries of the local random unitary matrices $U_n\in \mathcal U(d)$, and their complex conjugates, allows us to  replace the average over the local  unitary groups $\mathcal U(d)$ in Eq.~(\ref{eq:RandomMomentsQudits}) with an average over all elements of the respective Clifford groups $\mathcal C(d)$, yielding
\begin{equation}
\mathcal R_A^{(2)}=\frac{1}{|\mathcal C(d)|^{|A|}}\sum_{\alpha_1,\ldots,\alpha_k=1}^{|\mathcal C(d)|}  \expec{ \mathcal{O}^{(i_1)}_{U_{\alpha_1}} \cdot \ldots \cdot \mathcal{O}^{(i_k)}_{U_{\alpha_k}}}^2.
\label{eq:MomentUnitaryDesign}
\end{equation}
Equation~(\ref{eq:MomentUnitaryDesign}) is a general formula that allows one to calculate the second moment for general local dimension $d$ and arbitrary subsystems $A$. If one is interested in the specific case of systems of qubits, i.e., $d=2$, we can further use that the Clifford group has the property to map the multi-qubit Pauli group onto itself and thus each of the observables $\mathcal{O}^{(i)}_{U_{\alpha}}$ becomes equal to $\pm\sigma^{(i)}_{\alpha}$, where $\sigma^{(i)}_{\alpha}$ denotes the $\alpha$'th Pauli matrix acting on qubit $i$ of the total $N$-qubit system. All in all, 
Eq.~(\ref{eq:MomentUnitaryDesign}) thus simplifies to a sum over the squared elements of the correlations tensors $T^{(A)}_{\alpha_1,\ldots,\alpha_k}=\expec{\sigma^{(i_1)}_{\alpha_1} \ldots  \sigma^{(i_k)}_{\alpha_k}}$, leading to
\begin{align}
\mathcal R^{(2)}_A=\frac{1}{3^{|A|}} \sum_{\alpha_1,\ldots,\alpha_k=1}^{3} \big(T^{(A)}_{\alpha_1,\ldots,\alpha_k}\big)^2.
\label{eq:MomentSphericalDesign}
\end{align}  
Hence, in order to determine the moments $\mathcal R^{(2)}_A$ exactly it suffices to measure all elements of the correlation tensors $T^{(A)}$ which can be directly extracted from the full correlation tensor $T_{\alpha_1,\ldots,\alpha_N}=\expec{\sigma_{\alpha_1}\otimes \ldots \otimes \sigma_{\alpha_N}}$, with $\alpha_j=x,y,z$, which consists of $3^N$ elements. 

In conclusion, as the pure state concurrence~(\ref{eq:ConcurrenceMoments2}) and the corresponding lower bound (\ref{eq:ConcMixedMoments2}) are simple functions of the second moments $\mathcal R^{(2)}_A$, we can use Eq.~(\ref{eq:MomentSphericalDesign}) to measure them directly using $3^N$ measurement settings. Such a direct measurement, however, becomes impractical as soon as one reaches system sizes of several tens of qubits where, due to the exponential scaling $3^N$, the required number measurements becomes too large. In the latter regime it can be favourable to estimate the concurrence approximately using randomized measurements at the expense of a non-zero statistical error. In the next section we will analyse this statistical error and discuss the scaling of the required number of measurement settings, as well as the required number of projective measurements per individual measurement setting, in order to reach an estimate with a predefined accuracy.

\begin{figure*}[t!]
\begin{center}
\includegraphics[height=0.25\textwidth]{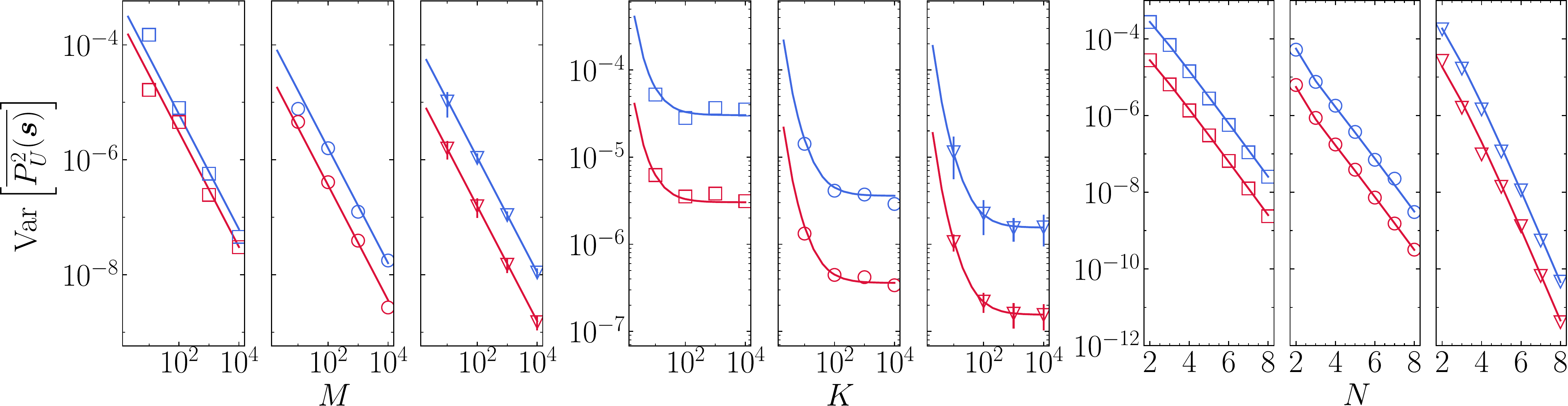}
\end{center}
\caption{
Left: Plot of the variance of the estimator of the squared population probabilities for $5$ qubits as a function of the number $M$ of sampled local measurement bases, with $K=10$ (upper blue circles) and $K=10^2$ (lower red circles). The left subplot corresponds to a random product state, the middle one to the GHZ state and the right subplot to a Haar random state. For the latter, the errorbar marks the standard deviation of 100 averaged Haar random states. Solid lines correspond to the analytical result obtained via an exact average with respect to the Haar measure. Middle: Plot of the same variacne as a function of the number of projective measurements $K$, with $M=10$ and $M=10^2$. Right: Plot of the same variance as a function of the number $N$ of qubits, with $M=10^2$ and $M=10^3$.
}
\label{fig_2a}
\end{figure*}

\section{Statistical estimation of the multiparticle concurrence}\label{sec:EstStatErr}

\subsection{Unbiased estimators of Eqs.~(\ref{eq:ConcPure}) and (\ref{eq:ConcMixed})}
In experiments randomized measurement protocols can only be realized with finite  samples of measurements. Consequently, an estimation of the randomized population probabilities contained in Eqs.~(\ref{eq:ConcPure}) and (\ref{eq:ConcMixed}), i.e., $\mathbb E_U\left[{P_{U}^2(\vec{s})}\right]$ and $\mathbb E_U\left[P_U(\mathbf{s})P_U(\mathbf{s}')\right]$, will involve a finite statistical error. Furthermore, in practice one also needs to estimate the population probabilities $P_U(\boldsymbol s)$ based on finitely many rounds of projective measurements.\\ 

\indent In the following we will assume that one round of randomized measurements consists of a sample of $M$ random measurement bases, each of which undergoes a finite number $K$ of projective measurements. The latter allow us to estimate the $P_U(\boldsymbol s)$, its second powers $P^2_U(\boldsymbol s)$, as well as cross-terms of the form $P_U(\boldsymbol s)P_U(\boldsymbol s')$ for each individual choice of random measurement bases defined by the local unitary transformation $U=U_1 \otimes \dots \otimes U_N$. In order to do so we first introduce an unbiased estimator of the population propability $P_U(\boldsymbol s)$ which reads $\widetilde{P}_U(\boldsymbol s)=Y(\boldsymbol s)/K$, where $Y(\boldsymbol s)$ denotes a random variable distributed according to the multinomial distribution defined by the distribution $\{P_U(\boldsymbol s)\}_{\boldsymbol{s}}$ with $K$ independent trials. Given $\widetilde{P}_U(\boldsymbol s)$ it is straightforward to derive appropriate estimators for its monomials (see App.~\ref{app:UnbEst} for details), yielding
\begin{align}
\widetilde{P}_U^{(2)}(\boldsymbol s)&=\frac{\widetilde{P}(\boldsymbol s)(K\widetilde{P}(\boldsymbol s)-1)}{K-1}, \\
\widetilde{P}_{U_i}^{(1,1)}(\boldsymbol s,\boldsymbol s')&=\frac{K}{K-1}\widetilde{P}(\boldsymbol s)\widetilde{P}(\boldsymbol s').
\end{align}
Given this set of unbiased estimators of the involved population propabilities we can go one step further and introduce statistically sound estimators of the relevant quantities contained in Eqs.~(\ref{eq:ConcPure}) and (\ref{eq:ConcMixed}), namely $\mathbb E_U\left[{P_{U}^2(\vec{s})}\right]$ and $\mathbb E_U\left[P_U(\mathbf{s})P_U(\mathbf{s}')\right]$, leading to 
\begin{align}
\overline{P^2_U(\mathbf{s})}
&=\frac{1}{M}\sum_{i=1}^M \widetilde{P}_{U_i}^{(2)}(\boldsymbol s),
\label{eq:EstPs2} \\
\overline{P_U(\mathbf{s})P_U(\mathbf{s}')} 
&=\frac{1}{M}\sum_{i=1}^M \widetilde{P}_{U_i}^{(1,1)}(\boldsymbol s,\boldsymbol s'). 
\label{eq:EstPsPsp}
\end{align}
 \begin{figure*}[t!]
\begin{center}
\includegraphics[height=0.25\textwidth]{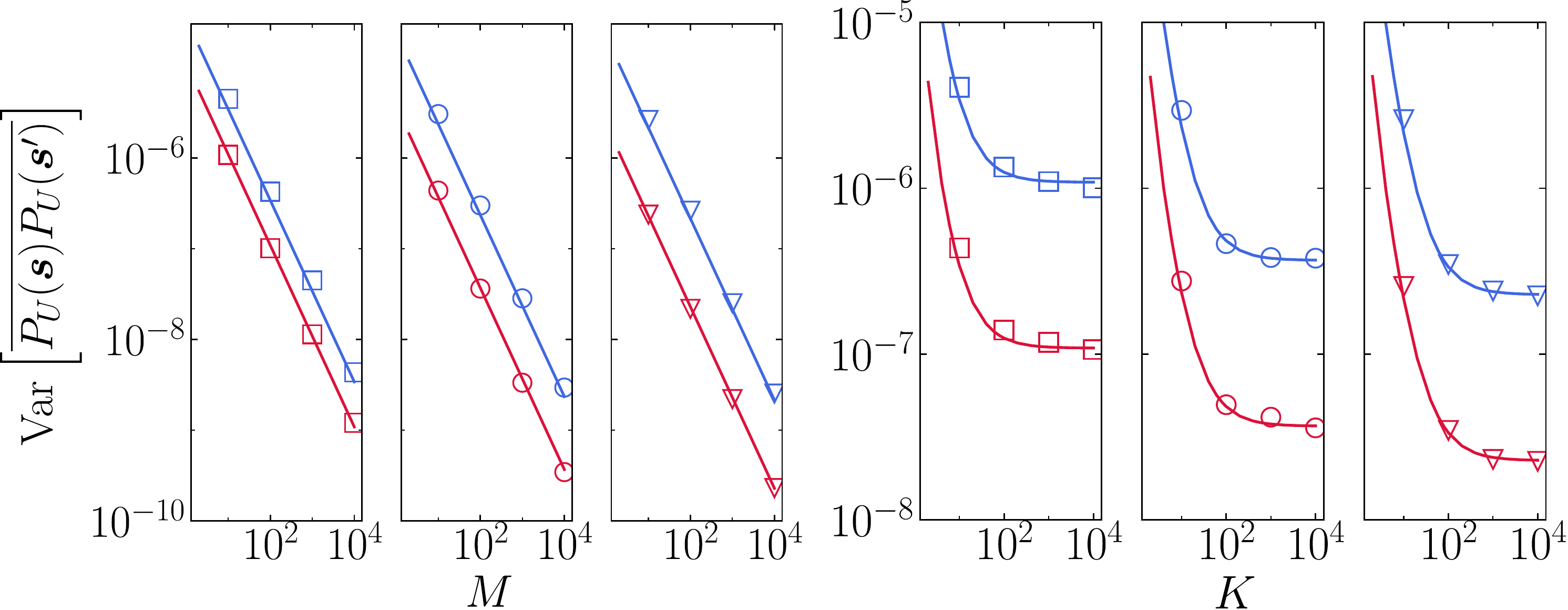}
\end{center}
\caption{
Left: Plot of the variance of the estimator~(\ref{eq:EstPsPsp}) for $5$ qubits averaged over all combinations of $\boldsymbol s$ and $\boldsymbol s'$, and as a function of the number $M$ of sampled local measurement bases, with $K=10$ (upper blue circles) and $K=10^2$ (lower red circles). The left subplot corresponds to a random product state, the middle one to the GHZ state and the right subplot to a Haar random state. Solid lines correspond to the analytical result obtained via an exact average with respect to the Haar measure. Right: Plot of the same variance as a function of the number of projective measurements $K$, with $M=10$ and $M=10^2$.
}
\label{fig_3}
\end{figure*}
\begin{figure}[b!]
\begin{center}
\includegraphics[width=0.4\textwidth]{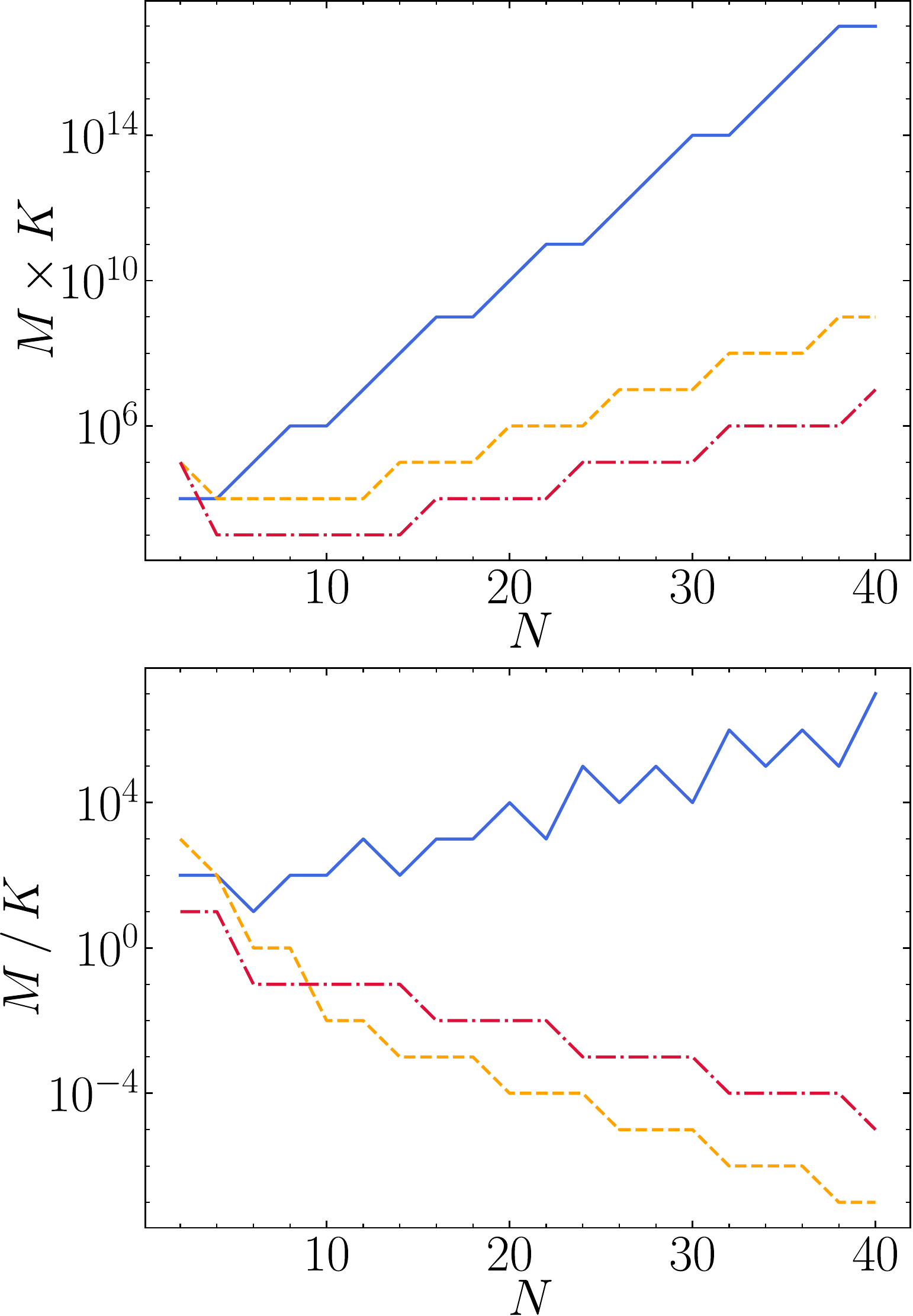}
\end{center}
\caption{Plot of the total number of measurements $M\times K$ (top) and the corresponding optimal ratio $M/K$ (bottom) required for an estimation of the concurrence with a relative error of at most $10\%$ as a function of the number of qubits for a GHZ state (solid blue line) and a Haar random state (dashed yellow line).  The dashed-dotted red line shows the respective results for the estimation the concurrence of a Haar random state using the estimator~(\ref{eq:EstCinclAvOfBitstrings}). 
}
\label{fig_4}
\end{figure}

The estimators~(\ref{eq:EstPs2}) and (\ref{eq:EstPsPsp}) thus reflect the fluctuations resulting from both  finite  $M$ and $K$ leading to a non-zero statistical error. In order to determine the latter different strategies can be pursued, one of which consists in a straightforward statistical approach exploiting the measurement data produced form either real or numerical experiments (see Fig.~\ref{fig_2a}). In this way it is possible to investigate the statistical error of the involved estimators for systems of limited size and for specific targeted quantum states~\cite{ZollerFirst,vermerschPRA,ZollerScience,ElbenPRA}, however, it is in general not possible to extrapolate the statistical effects to systems consisting of larger particle number. Alternatively, one can evaluate the statistical error analytically based on the given properties of the estimators' distributions; an approach that we will pursue further in Sec.~\ref{sec:StatErrs}.

Before moving on to the analysis of the statistical errors of Eq.~(\ref{eq:EstPs2}) and (\ref{eq:EstPsPsp}) we note that the squared averaged population probability $\mathbb E_U\left[P^2_U(\vec s)\right]$ does not longer depend on the bit-string $\vec s$. Hence, when estimating $\mathbb E_U\left[P^2_U(\vec s)\right]$ it can be advantageous to consider an unbiased estimator that includes also an average over the subsets $I\subset \{0,1\}^N$ of observed bit-strings $\boldsymbol s$. For instance, instead of Eq.~(\ref{eq:EstPs2}) we can use
\begin{align}
\tilde X=\frac{1}{|I|\times M}\sum_{\boldsymbol s\in I}\sum_{i=1}^M \widetilde{P}_{U_i}^{(2)}(\boldsymbol s),
\label{eq:EstCinclAvOfBitstrings}
\end{align}
which is also an unbiased estimator for the squared averaged population probability $ \mathbb E_U\left[{P_{U}^2(\vec{s})}\right]=\mathbb E_{\text{multi},U}[\tilde X]$. This procedure is particularly relevant for an increasing number of qubits $N$ in which case the probability of observing one particular bit-string can become vanishingly small. 

\subsection{Analysis of the statistical errors}\label{sec:StatErrs}

In order to analyse the statistical error of the estimators~(\ref{eq:EstPs2}), (\ref{eq:EstPsPsp}) and  (\ref{eq:EstCinclAvOfBitstrings}) we first have to evaluate their respective variances. For instance, in the case of Eq.~(\ref{eq:EstPs2}) this yields
\begin{align}
\text{Var}\left[\overline{P^{2}_{U}(\mathbf{s})}\right]=\frac{1}{M^2}\sum_{i=1}^M \text{Var}\left[\widetilde{P}_{U_i}^{(2)}(\boldsymbol s)\right],
\label{eq:VarEstPs2}
\end{align}
with
\begin{align}
\label{eq:VarEstPs2tilde}
\text{Var}\left[\widetilde{P}_{U_i}^{(2)}(\boldsymbol s)\right]&=\frac{1}{(K-1)^2}\Big\{(5-3 K) \mathbb E_U [P_U^4 (\boldsymbol s)]  \\
&+4 (K-2) \mathbb E_U [P_U^3 (\boldsymbol s)]+2 \mathbb E_U [P_U^2 (\boldsymbol s)]\Big\},  \nonumber
\end{align}
which shows that the underlying error depends on the higher-order randomized population probabilities $\mathbb E_U [P_U^2 (\boldsymbol s)]$, $\mathbb E_U [P_U^3 (\boldsymbol s)]$ and $\mathbb E_U [P_U^4 (\boldsymbol s)]$, which in turn are independent of the choice of the bit-string $\boldsymbol s$. In Fig.~\ref{fig_2a} we present a comparison between the numerically and analytically estimated values of the variance~(\ref{eq:VarEstPs2}) for a number of exemplary quantum states (see App.~\ref{app:UnbEst} for details on the calculations). We find that the variance decays for all states inversely with $M$, as expected form the central limit theorem, while its dependence on $K$ reaches a plateau after an initial decay. All in all we find that the numerical results agree very well with the analytical predictions also for varying number of qubits (see Fig.~\ref{fig_2a}~(right)).  

In Fig.~\ref{fig_3} we present  similar results for the estimator~\ref{eq:EstPsPsp}. Note that in this case the symbolic expression of the respective variance (see Eq.~(\ref{app:VarPsPsp}) in App.~\ref{app:UnbEst}) is more complicated as it depends on the cross-terms $\expi{P^t_U(\vec{s})P^k_U(\vec{s}')}$, with $t,k=1,2$, which themselves depend on the Hamming distance $D(\boldsymbol s,\boldsymbol s')$. For this reason, we calculated variance for all values of $\boldsymbol s$ and $\boldsymbol s'$, and present the average over the respective values in Fig.~\ref{fig_3}.  As for the variance of Eq.~(\ref{eq:EstPs2}) we find a good agreement with the analytical predictions for the states under consideration.   
 
Given the analytical expression of the variance~(\ref{eq:VarEstPs2}), it is straightforward to derive a confidence interval for the estimators~(\ref{eq:EstPs2}) and~(\ref{eq:EstPsPsp}) for the target state under consideration and a given number of measurements $M$ and $K$. To do so, we use the two-sided Cantelli  inequality (see Ref.~\cite{SchmidtSpringer2010}), yielding
\begin{align}
\text{Pr}[|\overline{P^2_U(\mathbf{s})}-\mathbb E[P^2_U(\mathbf{s})]|\geq \delta]\leq \frac{2{\text{Var}\big(\widetilde{P}_{U_i}^{(2)}(\boldsymbol s)\big)}}{\text{Var}\big(\widetilde{P}_{U_i}^{(2)}(\boldsymbol s)\big)+\delta^2},
\label{eq:eqcantelli}
\end{align}
which, by requiring that the confidence $1-\text{Prob}[|\tilde{\mathcal R}^{(t)}-{\mathcal R}^{(t)}|\geq \delta]$ of this estimation is at least $\gamma$, leads to the following minimal two-sided error bar: 
\begin{align}
\delta_\text{err} &= \sqrt{\frac{1+\gamma }{1-\gamma } \text{Var}\big(\widetilde{P}_{U_i}^{(2)}(\boldsymbol s)\big)}.
\label{eq:Precision1}
\end{align}

In order to do the final step of deriving a suitable error bar on the estimate of the concurrence we have to propagate the respective error~(\ref{eq:Precision1}) through the square root contained in the expression~(\ref{eq:ConcPure}). This can be done up to first order in $\delta_\text{err}$ using the standard rule for the propagation of uncertainties leading to
\begin{align}
\delta_C = \frac{\partial C}{\partial \mathbb E_U[{P_{U}^2(\vec{s})}]} \, \delta_\text{err} + \mathcal{O}\left(\delta_\text{err}^2\right).
\end{align}

Now we are in the position to determine the required number of measurements $M\times K$ as well as their ratio $M/K$ in order to estimate the concurrence up to an error of $\delta_C$. To do so, we fix a desired relative error of $10\%$ for the concurrence and determine the optimal values of $M$ and $K$ such that $\delta_C$ fulfils this error requirement. In practise, this is done by analytically evaluating $\delta_C$ for a number of values of $M$ and $K$ chosen from the list of values $10^1,\ldots,10^{16}$. In Fig.~\ref{fig_4} we present the results of this procedure as a function of the number $N$ of involved qubits. Note that this can be done efficiently thanks to the analytical estimates of the variance~(\ref{eq:VarEstPs2}) and, in principle, be carried out for an arbitrary number $N$ of qubits. Previous studies of this type focusing on estimations of the purity where bounded to values of $N\leq 10$ as they relied on numerical estimates of the underlying statistical errors~\cite{ZollerFirst,vermerschPRA,ZollerScience,ElbenPRA}. 

We also note that the exact analytical assessment of the statistical error of the estimator~(\ref{eq:EstCinclAvOfBitstrings}) is more involved because the sum over $\vec s$ leads to many cross-terms in the respective variance. However, in the particular case of Haar random states we can circumvent this problem because correlations between different bit-strings of Haar random states are with increasing $N$ exponentially suppressed~\cite{Ullah1946,Petz2004}. Making use of this fact allows us to evaluate the underlying variance and obtain an estimate of the respective statistical error (see App.~\ref{app:UnbEst} for details). As example we included the results of the latter calculation in Fig.~\ref{fig_3} showing that it can lead to an improvement as long as the dimension of the overall Hilbert space is small compared to the number $K$ of individual projective measurements per random measurement setting.

\section{Applications to typical multiparticle entangled states}\label{sec:ApplEntStates}

Having analyzed in detail the measurement resources required for the evaluation of Eqs.~(\ref{eq:ConcPure}) and (\ref{eq:ConcMixed}) in the last section we move on and investigate how the introduced randomized measurement protocol performs in practice. In this respect, we use the methods introduced in Sec.~\ref{sec:RandMeas} in order to characterize the multiparticle entanglement properties of examples of typical multiparticle entangled states and investigate the observed  performance in the presence of noise in form of gate errors. 

\subsection{Analytical results for pure states}\label{sec:ConcAnaVal} 
To begin with, we will summarize some important analytical expressions of the concurrence for several examples of multiparticle states and discuss their respective asymptotic behavior in the limit of large particle numbers. 
We note that the analytical expressions for the concurrence of pure GHZ-states of $N$ qubits can be easily derived from the fact that all its reduced states are maximally mixed states of rank $2$ and thus have a purity of $1/2$~\cite{Carvalho2004,Mintert2005a}, leading to:
\begin{align}
C_N\big(\ket{\text{GHZ}_N}\big)=2^{1-\frac{N}{2}} \sqrt{2^{N-1}-1},
\label{eq:GHZanalytical}
\end{align}
which yields $\sqrt 2$ in the limit $N\rightarrow \infty$. 

\begin{figure}[t!]
\begin{center}
\includegraphics[width=0.48\textwidth]{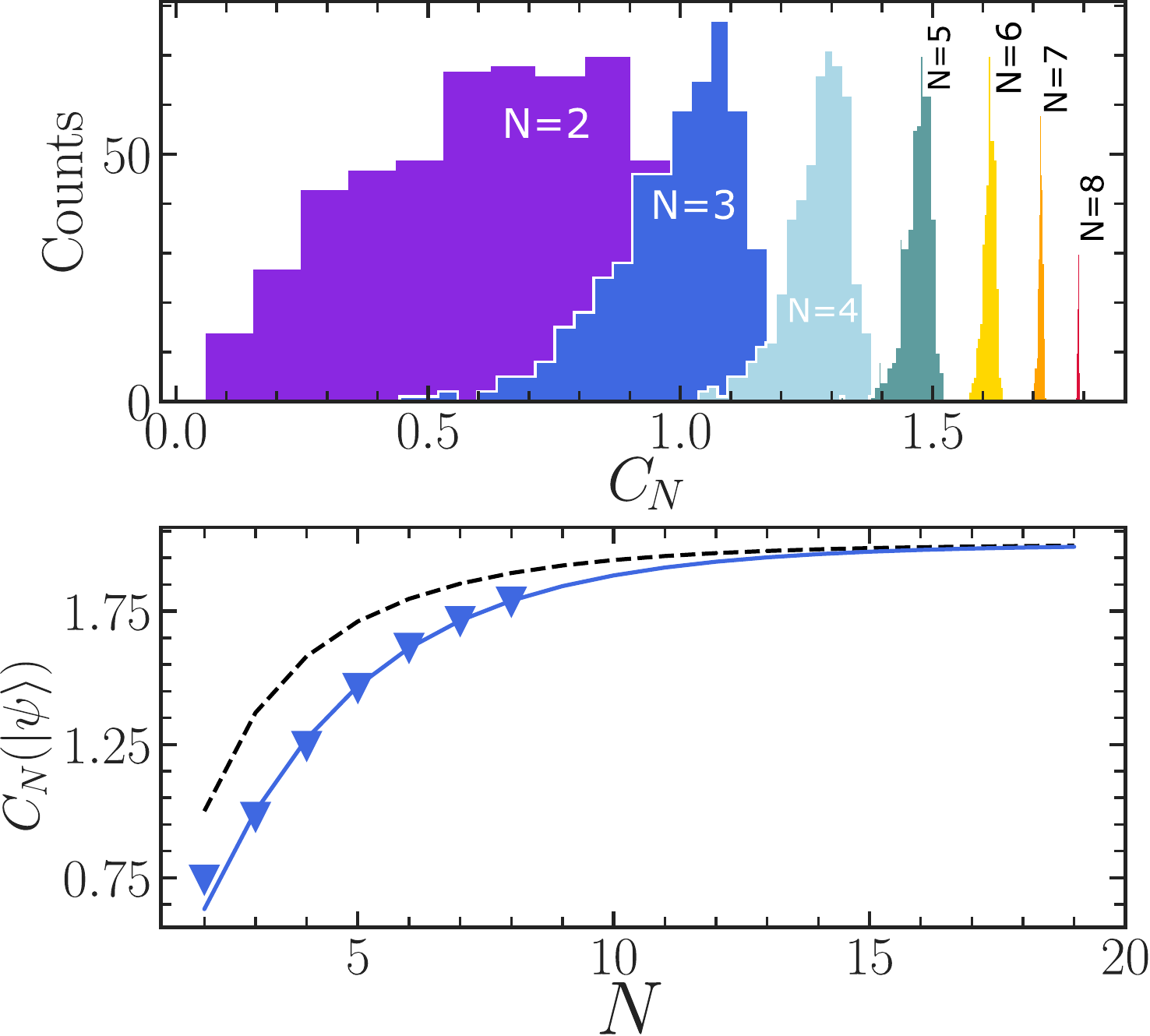}
\end{center}
\caption{Top: Histograms of the values of the multiparticle concurrence~(\ref{eq:concurrence}) evaluated for samples of $10^3$ Haar random states for $N=2$ (violet, left) to $N=8$ (red, right) qubits. Bottom: Plot of the mean values associated to the distributions presented above as a function of the number of qubits $N$ (blue triangles). As comparison the analytical law~(\ref{eq:ConcHaarAna}) for the mean value of the concurrence for Haar random states is shown (solid blue line), as well as the overall upper bound of the maximum of the concurrence presented in Eq.~(\ref{eq:ConcUpperBound}) (black dashed line). 
}
\label{fig_5}
\end{figure}

Furthermore, we derive in the following also an analytical expression for the concurrence of Haar random pure states of $N$ qubits. In order to do so we first note that a pure Haar random state reads

 $\ket{\psi}=U\ket{0}^{\otimes N}$, with $U\in \mathcal U(2^N)$ picked uniformly according to the Haar measure. Hence, the resulting concurrence of the output state is a polynomial functions of the unitary transformation $U$ whose average over the unitary group can be evaluated using the well-known expression
\begin{align}
&\int_{\mathcal{U}(d)} U_{i_1,j_1}\dots U_{i_t,j_t}U^*_{\tilde i_1,\tilde j_1}\dots U^*_{\tilde i_t,\tilde j_t}dU \label{eq:HaarIntU} \\
= &\sum_{\pi, \sigma \in S_t} \delta_{i_1, {\tilde i}_{\pi(1)}}\dots \delta_{i_t, \tilde i_{\pi(t)}}\delta_{j_1, \tilde j_{\sigma(1)}}\dots \delta_{j_t, \tilde j_{\sigma(t)}} \, \text{Wg}_d(\pi^{-1}\sigma),
\nonumber
\end{align}
where the sum runs over the elements of the symmetric group $S_t$ and  $\text{Wg}_d$ denote the so-called Weingarten functions which depend on the structure of the permutation $\pi^{-1}\sigma$ and the dimension $d$~\cite{HaarAverages,Weingarten}. Doing so with the square of the concurrence~(\ref{eq:ConcPure}) leads to the following expression
\begin{align}
\exppsi{C_N(\psi)^2} 
&=4-\frac{8(1+d)^N}{2^N(1+d^N)},
\label{eq:ConcHaarAna}
\end{align}
where we used the notation $\exppsi{\ldots}$ to denote the analytical average over $U$ and thus over the Haar random state $\ket\psi$. We emphasize that the square root of Eq.~(\ref{eq:ConcHaarAna}) provides also a good approximation of the average $\exppsi{C_N(\psi)}$ already for moderate numbers of $N$ due to the concentration of the of concurrence around its mean value (see Fig.~\ref{fig_5}). The latter is a direct consequence of the concentration of measure phenomenon occurring for samples of Haar random quantum states in Hilbert space of growing dimension~\cite{HaarConcentration1,HaarConcentration2}. 

Using Eq.~(\ref{eq:ConcHaarAna}) we find that the average concurrence for Haar random multiqubit states converges to $2$ if the number of qubits $N$ goes to infinity. Hence, while GHZ states yield a larger concurrence for small qubit numbers, i.e., $N=2,3,4$, the concurrence of Haar random states generally increases faster for large qubit numbers and finally also reaches a larger asymptotic value $C_\infty(\ket{\text{Haar}_\infty})>C_\infty(\ket{\text{GHZ}_\infty})$. Lastly, it remains the question whether $2$ is also the global maximum of the concurrence in the limit $N\rightarrow \infty$? To answer this question we make the hypothetical assumption that all subsystems of a pure $N$ qudits state are maximally mixed 
and thus all the corresponding purities contained in Eq.~(\ref{eq:concurrence}) become minimal. Hence,  we  find that all the purities are equal to $1/d_A$, where $d_A$ denotes the dimension of the respective subsystem under consideration. In the case of a system of qubits we thus have $d_A=2^{|A|}$ and summing over all possible subsystems leads to the formula 
\begin{align}
2 \sqrt{1-\frac{1}{2^N}-{\sum _{k=0}^{N-1} \frac{\binom{N}{k}}{2^{N+k}}}}=2^{1-N} \sqrt{1+4^{N}-2^{N}-3^{N}},
\label{eq:ConcUpperBound}
\end{align}
which provides an upper bound of the global maximum of the concurrence~(\ref{eq:concurrence}). Note that the  assymptotic value of Eq.~(\ref{eq:ConcUpperBound}) in the limit $N \rightarrow \infty$ is also $2$, while for each finite $N$ it  is strictly larger than Eq.~(\ref{eq:ConcHaarAna}). 

\subsection{Influence of noise on multiparticle entanglement}\label{sec:ApplHaarGHZNoise}

\begin{figure}[b!]
\begin{center}
\includegraphics[width=0.48\textwidth]{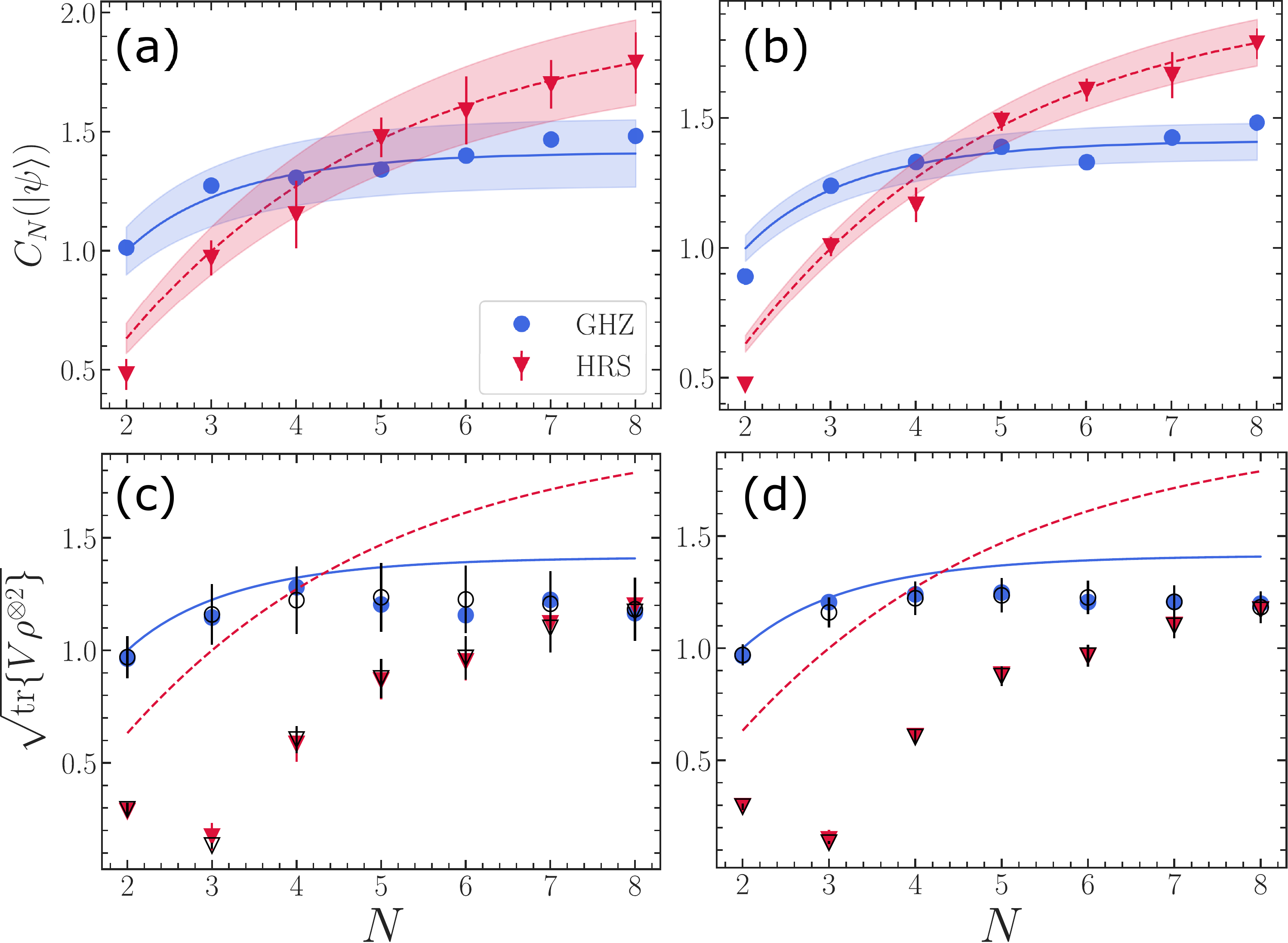}
\end{center}
\caption{
Numerical estimation of the multiparticle concurrence~(\ref{eq:ConcPure}) (a,b) and its lower bound~(\ref{eq:ConcMixed}) (c,d) of GHZ-states (blue circles) and Haar random states (red triangles) as a function of the number of qubits $N$. The values of $M$ and $K$ are chosen in order to reach a relative error of $10\%$ (a) and $5\%$ (b) according to the analysis presented in Sec.~\ref{sec:EstStatErr}. While solid and dashed lines present the corresponding analytical predictions for pure states, black circles and triangles indicate the respective true values of the concurrence's lower bound obtained via Eq.~(\ref{eq:ConcLowerBound}). Note that red triangles and the corresponding error bars have been obtained samples of $30$ Haar random states. 
Plots (c) and (d) present the noise-prone cases where the states are produced by a quantum circuits consisting of single- and two-qubit gates inflicted with local depolarizing errors of $0.01\%$ and $0.1\%$, respectively. In order to approximate mixed Haar random states we used random quantum circuits containing $500$ randomly sampled gates from a universal gate set (see Fig.~\ref{fig_1} and Sec.~\ref{sec:RandCirc}). 
}
\label{fig_6}
\end{figure}

Given the statistical analysis of the required measurement resources for an estimation of the multiparticle concurrence~(\ref{eq:ConcPure}), we are now in the position to apply these insights in practice. We first do so in the case of  pure GHZ as well as Haar random states and compare the results to the analytical expressions presented in the previous Sec.~\ref{sec:ConcAnaVal}. The two plots in the upper panel of Fig.~\ref{fig_6} present numerical estimates of the concurrence with the measurement resources $M$ and $K$ chosen in such a way that the resulting statistical errors remains below $10\%$ and, respectively, $5\%$ of the absolute value of the concurrence. While the fluctuations of the resulting estimates are apparent one clearly observes that they remain below the anticipated relative error bounds of $10\%$ and $5\%$. 

In a further analysis we estimate the lower bound of the multiparticle concurrence of noisy versions of the respective pure states with randomized measurements using Eq~(\ref{eq:ConcMixed}). In order to stay close to experimental implementations using NISQ devices, we produce the respective GHZ and Haar random states with simulated quantum circuits consisting of series' of single- and two-qubit gates for which we assume local depolarizing errors with error-propabilities $\varepsilon_1$ and $\varepsilon_2$, respectively. However, if the latter are chosen too large the resulting output states will be considerably mixed and the associated lower bound of the concurrence very small or even zero. 
Hence, in order to produce states with a reasonable fidelity, i.e., such that the lower bound of the concurrence is above zero, the magnitude of  $\varepsilon_1$ and $\varepsilon_2$ should not be to large. We can roughly estimate the resulting fidelities of the final output states by summing up the effects of all single- and two-qubit gate errors as $\epsilon=1-(1-\varepsilon_1)^{\#_{1\text{-qu}}}(1-\varepsilon_2)^{\#_{2\text{-qu}}}$, where ${\#}_{1\text{-qu}}$ and ${\#}_{2\text{-qu}}$ denote the total numbers of applied single- and two-qubit gates, respectively. The estimated overall accumulated error when producing GHZ states of $N$ qubits is thus given by $1-(1-\varepsilon_1)(1-\varepsilon_2)^{N-1}$ as it requires the application of exactly $N-1$ CNOT gates and only a single Hadamard gate. Hence, the number of required two-qubit gates grows linearly with the number of involved qubits and the overall accumulated error remains below $10\%$ even for two-qubit gate errors of about $1\%$. 

 The Haar random states, however, can only be approximated by a series of gates that are chosen randomly from a universal set of gates (in Sec.~\ref{sec:RandCircuits} we discuss one possible way of doing so) and thus one has to find a tradeoff between the required randomness one wants to achieve and the total error inflicted by the executed gate operations. If we assume that the latter circuits consist of overall $n_\text{gates}$, with twice as many single than two qubit gates, we accumulate an overall error of  $1-(1-\varepsilon_1)^{n_\text{gates}2/3}(1-\varepsilon_2)^{n_\text{gates}/3}$. Furthermore, we need to apply at least $n_\text{gates}>500$ gates in order to reach a sufficient amount of randomness in the case of $N=9$ qubits. Taking these competing factors into account we estimate that for the errors $\varepsilon_1=0.01\%$ and $\varepsilon_2=0.1\%$ the overall accumulated error does not exceed $20\%$ and thus a reasonable fidelity of the respective Haar random state is reached.  
 
 We thus simulated the respective circuits with the above error rates and and estimated the concurrence using Eq.~(\ref{eq:ConcMixed}). The results are presented in Fig.~\ref{fig_6}(c) and (d). Note that for the simulation of Eq.~(\ref{eq:ConcMixed}) in Fig.~\ref{fig_6}(c) and (d) we used the same measurement numbers $M$ and $K$ that have been used for the respective pure states in order to reach a relative  error of $10\%$ and $5\%$, respectively. Even though this is only a rough estimate the obtained results agree well with the exact values of the concurrence's lower bound~(\ref{eq:ConcLowerBound}) which are also depicted in Fig.~\ref{fig_6}. Motivated by this result we move on and apply the respective protocol for estimating the concurrence other multiparticle entangled states produced by different classes of random quantum circuits. 

\section{Applications to the characterization of random quantum circuits}\label{sec:RandCirc}

\subsection{Random quantum circuits}\label{sec:RandCircuits}
\begin{figure}[t!]
\begin{center}
\includegraphics[width=0.48\textwidth]{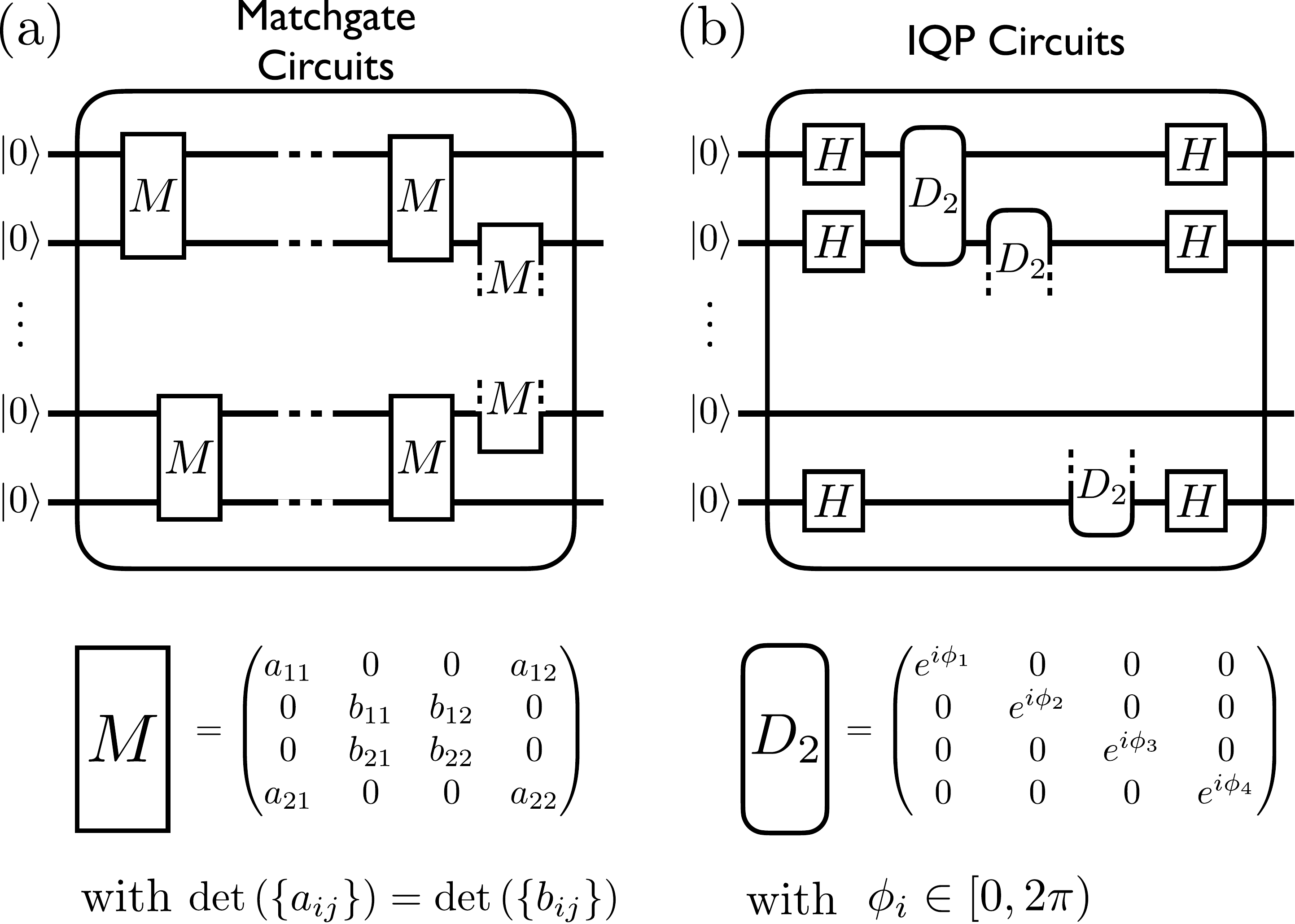}

\hspace{0.3mm}

	\begin{tabular}{|l|l|l|l|}
		\hline
		\textbf{Gate Set} & \textbf{Universality}  & \textbf{Classical Simulatability}\\
		\hline
		\hline
		 $I_\text{Uni}$ & Yes 
		& \textit{Never}~\cite{NielsenChuang}\\
		\hline
		$I_\text{MG}$ & No & \textit{One, Strong}~\cite{MG_1,MG_2} \\ 
		\hline
		$I_\text{IQP2}$ & No  & \textit{Many, Weak}~\cite{IQP3} \\ 
		\hline
	\end{tabular}
\end{center}

\caption{Representation of matchgate (left) and IQP (right) circuits together with the matrix representation of the native gate operations $M$ and $D_2$, respectively. The table summarizes the properties of the considered types of random quantum circuits. The second column indicates  whether the respective gate set gives rise to universal quantum computations. The third column summarizes the complexity of the circuits showing whether or not the circuit is classically simulatable and under which conditions. Strong and Weak indicate whether it is possible to classically simulate the circuits output probabilities or to classically sample from it, respectively. \textit{One} and \textit{Many} say whether the task involves a single qubit or many qubits.
}
\label{fig_7}
\end{figure}

In the following we consider quantum circuits which are defined through sequences of unitary gates that are drawn randomly from a predetermined gate set $\mathcal I$ and applied to randomly selected subsets of qubits (see Fig.~\ref{fig_1} and \ref{fig_7}). In particular, we prepare the initial state of the $N$ qubits in the ground state $\ket{0}^{\otimes N}$ and apply exactly $T$ randomly drawn gates, i.e.,  $T$ can be considered as a discrete time parameter which also denotes the total count of quantum gates that have been applied. Note that selection of gates form the set $\mathcal I$ as well as the choice of qubits to which they are applied is entirely random. The $N$ qubit output state of such a random quantum circuits consequently depends on the number of applied gate operations $T$, and the properties of the gate set $\mathcal I$ under consideration. We will regard three distinct types of gate sets which have fundamentally different properties concerning their universality and classical simulability, and study the entanglement that is produced by them.

With the goal of performing universal quantum computation in mind one usually considers universal gate sets, i.e., sets which allow to approximate any $N$ qubit unitary transformation with arbitrary precision $\epsilon$ \cite{NielsenChuang}. One of the most famous universal gate sets consist of the two-qubit controlled not gate $C_X$, the Hadamard gate $H$ and the $T=\exp{[-i\sigma_z \pi/8]}$ gate, and we refer to it in the following as $\mathcal{I}_{\text{uni}}$~\cite{NielsenChuang} (see Fig.~\ref{fig_1}). A random quantum circuit consisting of gates form $\mathcal{I}_{\text{uni}}$ is expected to approximate an overall Haar random unitary transformation over $N$ qubits once a threshold time $T^*$ is reached. The universal gate set $\mathcal{I}_{\text{uni}}$ has also been used to approximate Haar random states in Sec.~\ref{sec:ApplEntStates}. Also note that noisy variants of such universal random circuits are at the heart of the first demonstrations of quantum computational advantages based on cross-entropy benchmarking~\cite{QuSup}.

In the following we will investigate the entanglement properties of other, so-called restricted, classes of quantum circuits, i.e., circuits produced by gate sets that are in general not universal. 
A famous example of such a restricted class of quantum circuits is that consisting of so-called Clifford transformations which are generated by the set $\mathcal I_\text{Clif}=\{C_X,H,P=\exp{[-i\sigma_z \pi/4]}\}$. The latter are, by virtue of the Gottesman-Knill theorem, always classically simulatable when initiated in the state $\ket{0}^{\otimes N}$ and readout in the computational basis~\cite{NielsenChuang}. 
Another class of quantum circuits that is known to be efficiently simulatable are so-called nearest-neighbour matchgate circuits (see Fig.~\ref{fig_7}(a)). Matchgates are two-qubit gates that consist of two single qubit gates with equal determinant that act on the even and odd parity subspace of the two qubits, respectively~\cite{MG_1,MG_2}.  
Matchgate circuits on $N$ qubits have also been shown to be equivalent to a  system of non-interacting fermions in one dimension which is governed by  interactions of at most quadratic order in the fermion creation and annihilation operators~\cite{TerhalDivincenzo2002}. In the random circuit model we generate in each time step a random matchgate and apply it to a random pair of nearest-neighbour qubits. Lifting the nearest-neighbour restriction of matchgates circuits promotes them to the realm of universal for quantum computation~\cite{MG_1}.

\begin{figure}[t!]
\begin{center}
\includegraphics[width=0.48\textwidth]{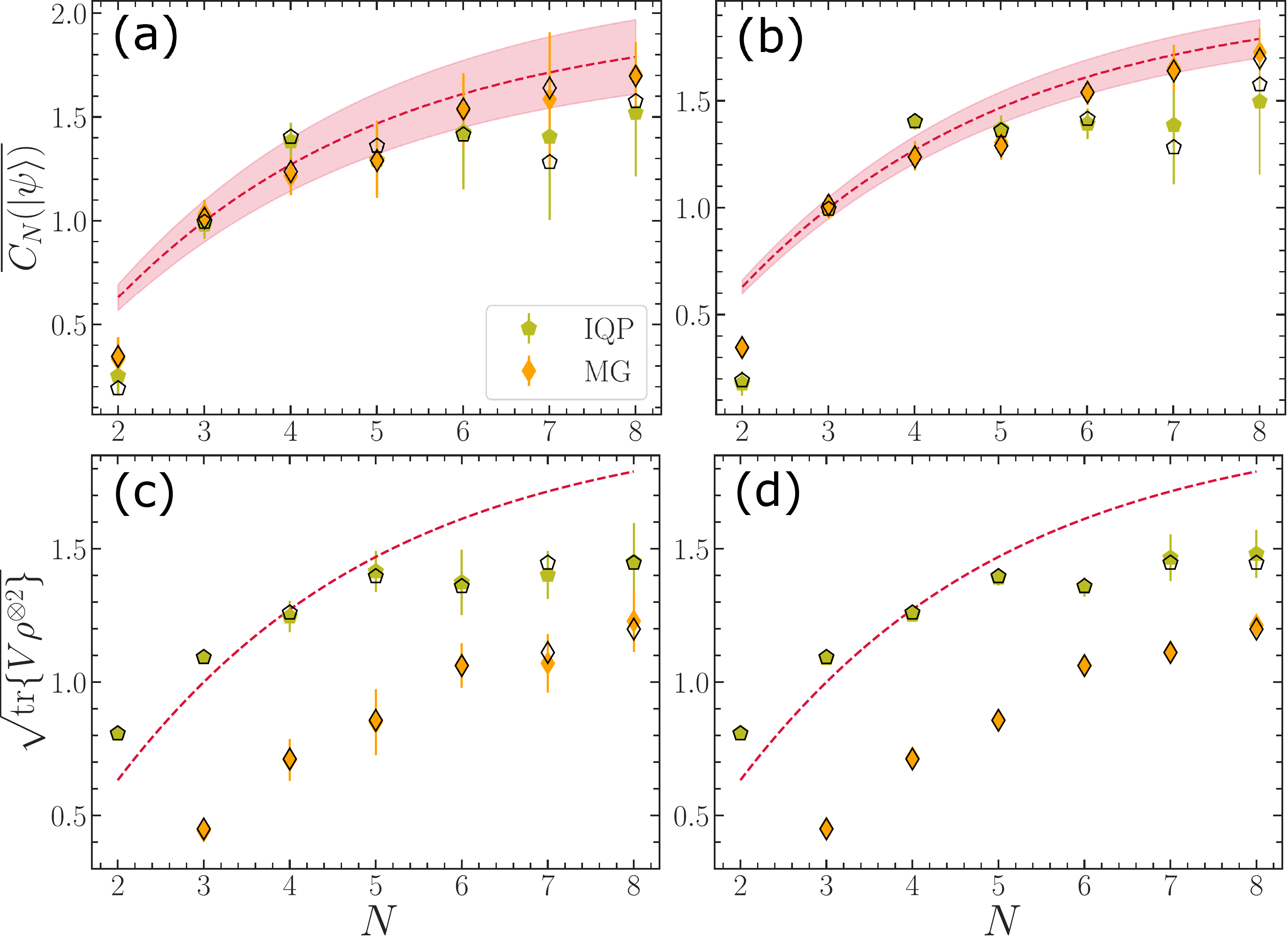}
\end{center}
\caption{Numerical estimation of the multiparticle concurrence~(\ref{eq:ConcPure}) (a,b) and its lower bound~(\ref{eq:ConcMixed}) (c,d) of output states of random IQP (green pentagons) and MG (orange diamonds) circuits. Plot (a) and (b) show the noiseless cases with $M$ and $K$ chosen in order to reach a relative error of $10\%$ (a,c) and $5\%$ (b,d), respectively, according to the analysis for Haar random states presented in Sec.~\ref{sec:EstStatErr}. IQP circuits consist of $N(N-1)/2$ diagonal gates and MG circuits of $150$ randomly sampled nearest-neighbor match-gates. Error bars are obtained by resampling the respective states $30$ times. The lower plots (c) and (d) present the cases where we assumed that each two-qubit gate used to produce the respective states has a local depolarizing error of $0.1\%$. Black symbols indicate the respective true values of the concurrence's lower bound obtained via Eq.~(\ref{eq:ConcLowerBound}). 
}
\label{fig_8}
\end{figure}

Lastly, we consider the class of commuting quantum circuits which are made up of gates diagonal in the computational basis (see Fig.~\ref{fig_8}(b)). The latter become non-trivial if the qubits are initiated and readout in local bases that are orthogonal to the computational bases, e.g., $\{\ket{x_i}^{\otimes N}\}_{i=1}^N$, with $x_i=\pm$. Due to the commuting property of the diagonal gates there is not a natural time ordering of gates for a given circuit and thus the resulting class of circuits is referred to as instantaneous quantum polynomial-time (IQP) circuits. IQP circuits do not allow for universal quantum computation but they are  known to be in general hard to simulate classically~\cite{IQP1,IQP2}. 
Specific designs of IQP circuits deal with diagonal gates of the form $W_r=\text{diag}\{e^{i\phi_1},\ldots, e^{i\phi_{2^r}}\}$, with independent and uniformly sampled $\phi_i\in[0,2\pi)$, which act on a subset of $r$ qubits and are applied to all combinations of $r$ qubits in  random order~\cite{PhaseRandomCircIQP}. In particular, we will consider IQP circuits with $r=2$ which consist of precisely $N(N-1)/2$ diagonal gates. For the former type, i.e. $r=2$, a restriction to gates acting only on nearest-neighbor qubits once again makes the cicuits classically simulatable~\cite{IQP3}.

\subsection{Numerical results}

As a last application we use three of the aforementioned random circuits to test our randomized measurement protocols on further examples of mutliparticle entangled states. So far, we have considered random circuits consisting of the universal gate set $I_\text{Uni}=\{H,T,C_X\}$ in order to approximate Haar random states (see Sec.~\ref{sec:ApplHaarGHZNoise}). Here we focus on two examples of random circuits produced by restricted gates sets, i.e. nearest-neighbor matchgates and IQP circuits (see Fig.~\ref{fig_7}). The corresponding gate set $I_\text{MG}$ and $I_\text{IQP}$ are non-universal and also classically simulatable, however, in some cases can produce more entanglement than the aforementioned universal set $I_\text{Uni}$ for which reason they provide an appropriate test case for our randomized measurement protocol.

We start by the noiseless cases presented in Fig.~\ref{fig_8}(a) and (b) where we estimated the concurrence of samples of output states of the above mentioned random circuits. To do so, we used in all three cases the same measurement resources $M$ and $K$ as determined for Haar random states of the respectively same number of qubits. We find that even though $I_\text{MG}$ and $I_\text{IQP}$ produce very different ensembles of states the resulting concurrence agrees well, i.e., within one standard deviation,
with the respective analytically determined values using Eq.~(\ref{eq:concurrence}). 
However, we find that $I_\text{MG}$ and $I_\text{IQP}$ circuits lead in general to larger fluctuations of the concurrence as compared to $I_\text{Uni}$. Note that the former have been analysed by resampling the concurrence of the corresponding random circuit $100$ times in order to numerically determine the underlying standard deviation, as shown in Fig.~\ref{fig_8}. 

Finally, we consider the same random circuits but with error-prone gates simulated by additional local depolarizing channels with the same single- and two-qubit error probabilities as used also in Sec.~\ref{sec:ApplHaarGHZNoise}, i.e. $\varepsilon_1=0.01\%$ and $\varepsilon_1=0.1\%$. The results for estimating the concurrence's lower bound of the respective output states are presented in Fig.~\ref{fig_8}(c) and (d). Again we find that using the same measurement resources as for the corresponding noiseless cases one achieves a good agreement with the corresponding exact values. However, in comparison to the noiseless cases the resulting estimates do not fluctuate stronger which might be explained by the increased mixing of local subsystems. The more the one qubit reduced states are mixed the less they fluctuate over the local unitary ensembles while performing randomized measrurements on them. Hence, this indicates that protocols based on randomized measurements might be a relevant candidates for characterization of entanglement in close to random, noisy quantum circuits in the intermediate regime.   

\section{Conclusions}\label{sec:Conclusion}
In this article we investigated how to estimate the entanglement content, as measured by the multiparticle concurrence, of many-body quantum systems employing protocols based on randomized measruements. We formulated schemes for measuring the multiparticle concurrence of pure states as well as a corresponding lower bound in the case of mixed states. Further on, we analyzed in detail the occurring statistical error when estimating the involved quantities with appropriate unbiased estimators and derrived exact scaling laws of the required measurement resources for estimating the concurrence of an important subclass of multiparticle entangled states. 

We demonstrated the introduced protocols by numerically analysing the multiparticle concurrence of the aforementioned class of quantum states, as well as for ensembles of output states of different classes of random quantum circuits, such as  matchgate and IQP circuits. Finally, we investigated the influence of noise in terms of single- and two-qubit gate errors on the multiparticle entanglement of the states under consideration and thereby showed that the obtained results on the required measurement resources prove useful in the noisy intermediate-scale regime. However, we also show that the required measurement resources strongly depend on the underlying class of quantum states under consideration showing that a detailed statistical analyses was justified.  

All in all, the outlined randomized measurement protocols are promising tools for the analysis of multiparticle entanglement in NISQ devices. Nevertheless, the measurement resources required for an estimation of the concurrence with a reasonably small statistical error increase quickly when reaching regimes of large particles numbers, i.e. beyond $N\approx 30$, rendering the presented protocols impractical. Hence, while moving towards larger and larger particle numbers of NISQ devices with improved quality one has to develop alternative tools that deal with the aforementioned problem. A possible solution in this direction is to exploit more information about the actual quantum states under investigation, e.g., the fact that they are likely contained in a subspace of limited qubit excitations or entanglement content. Alternatively, one might employ more involved, non-local measurement schemes on an extended Hilbert space~\cite{Mintert2005a,AolitaConc2}, in order to analyze its multiparticle entanglement in a more efficient manner.

\begin{acknowledgements}
We are indebted to Andreas Buchleitner, Fernando de Melo, Nikolai Wyderka, Otfried G\"uhne, Satoya Imai  and Xiao-Dong Yu for fruitful discussions. We acknowledge support by the state of Baden-Wuerttemberg through bwHPC.  AK acknowledges support from the Georg H. Endress foundation.

\end{acknowledgements}

\appendix
\section{Multiparticle concurrence from randomized measurements}
\subsection{Derivation of Eq.~(\ref{eq:ConcPure}) and~(\ref{eq:ConcMixed})}\label{app:A1}
The concurrence of a pure $N$ qudit state $\ket\psi \in \mathbb{C}^d$ is defined as~\cite{Carvalho2004,Mintert2005a}
\begin{equation}
C_N(\ket\psi) = 2\sqrt{1 - \frac{1}{2^N}\sum_{A \subseteq \{1,\ldots,N\}} \tr{( \varrho_A^2)}},
\label{app:concurrence}
\end{equation}
where $\varrho_A=\text{tr}_{\bar{A}}[\ketbra{\psi}{\psi}]$, with $\bar{A}=\{1,\ldots,N\} \setminus A$, denotes the reduced density matrix of the pure state $\ket\psi$ with respect to the subsystem associated to the subset $A\in\{1,\ldots,N\}$. Note that the sum in Eq.~(\ref{app:concurrence}) runs over all subsets including the empty set for which we have $\varrho_\emptyset=1$. In the following we will show that one can evaluate Eq.~(\ref{app:concurrence}) using locally randomized measurements. To do so, we regard the  population probabilities 
\begin{align}
P_{U}(\vec{s}) = \tr\left\{U\varrho U^\dagger \ketbra{\vec{s}}{\vec{s}}\right\},
\label{app:PopProb}
\end{align}
where $\ket{\vec{s}} = \ket{s_1,\dots, s_N}$, with $s_i = 1\dots d$, denotes an arbitrary element of the computational basis of $N$ qudits and $U = U_1 \otimes \dots \otimes U_N$, with $U_i \in \mathcal{U}(d)$, a randomly drawn local unitary transformation. Further on, upon averaging the square of Eq.~(\ref{app:PopProb}) over the $U$'s with respect to the local Haar measure on each of the individual qudit subspaces we find
\begin{align}
\mathbb E_U\left[{{P_U(\vec{s})^2}}\right] &=\mathbb E_U\left[{\tr\left\{ \varrho U \ketbra{\vec{s}}{\vec{s}}U^\dagger \right\}^2}\right]\nonumber\\
&= {\tr\left\{\varrho^{\otimes 2}\mathbb E_U\left[{\left(U \ketbra{\vec{s}}{\vec{s}}U^\dagger\right)^{\otimes 2}}\right]\right\}}\nonumber\\
&= \tr\left\{\varrho^{\otimes 2}\bigotimes_{i=1}^N\mathbb E_{U_i}\left[{U_i^{\otimes 2}\ketbra{{s_i}}{{s_i}}^{\otimes 2}{U_i^\dagger}^{\otimes 2}}\right] \right\}\nonumber\\
&= \left(\frac{D_d^{(2)}}{2}\right)^N \tr \left\{\varrho^{\otimes 2}(P_+ \otimes \ldots \otimes P_+) \right\}.
\label{eq:popprob}
\end{align}
In the last line of Eq.~(\ref{eq:popprob}) we used for $t=2$ the following relation 
\begin{align}
\mathbb E_{ U(d)}\left[{U^{\otimes t}\ketbra{s}{s}^{\otimes t}{U^\dagger}^{\otimes t}}\right]
= D^{(t)}_d  P_+,
\label{app:ProjectorSymt}
\end{align}
where $P_+$ denotes the projector on the symmetric subspace of $({\mathbb C^d})^{\otimes t}$ and $D^{(t)}_d =t!(d-1)!/(t+d-1)!$ the inverse of its dimension. For $t=2$, we can write $P_+=(\mathbb 1+S)/2$, where $S$ denotes the swap operator on $({\mathbb C^d})^{\otimes 2}$. Using the latter and the fact that $\text{tr}[S\varrho^{\otimes 2}]=\text{tr}[\varrho^2]$ in Eq.~(\ref{eq:popprob}) allows us to arrive at the expression 
\begin{align}
\overline{{P(\vec{s})^2}} &= \left(\frac{D_d^{(2)}}{2}\right)^N \tr \left\{\prod_{i=1}^{N}(S_{i} + \mathbb{I})\, \varrho^{\otimes 2}\right\}  \nonumber \\
&= \left(\frac{D_d^{(2)}}{2}\right)^N \sum_\alpha \tr \varrho_\alpha^2
\end{align}
and thus shows that the concurrence~(\ref{app:concurrence}) can be expressed as
\begin{equation}
C_N(\ket\psi) = 2\sqrt{1 -\frac{d^N(d+1)^N}{4^N}\sum_{\vec s \in \{0,1\}^N} \mathbb E_U\left[{P_{U}^2(\vec{s})}\right]}.
\label{eq:concpopprob}
\end{equation}

Further on, we note that the lower bound of the multiparticle concurrence for mixed states (given in Eq.~(\ref{eq:ConcLowerBound}) of the main text) can be expressed as a combination of purities evaluated on subsystems $A\subset \{1,\ldots,N\}$ of the $N$-party space, as follows
\begin{align}
C(\varrho) \geq 2^{1-\frac{N}{2}}\sqrt{1- \sum_{A\subseteq \{1,\ldots,N\}}\text{tr}[\varrho_A^2]   + (4-2^{2-N})\tr(\varrho^2)}.
\label{app:ConcLowerBound}
\end{align}
Hence, using Eq.~(\ref{eq:concpopprob}) together with  the the well-known formula for the purity in terms of randomized measurements~\cite{ZollerScience}, i.e., 
\begin{align}
\tr[\varrho^2]=d^{N} \sum_{\mathbf{s},\mathbf{s}'} (-d)^{-D(\mathbf{s}, \mathbf{s}')} \overline{P_U(\mathbf{s})P_U(\mathbf{s}')},
\end{align}
where $D(\mathbf{s}, \mathbf{s}')$ denotes the Hamming distance (as explained after Eq.~(\ref{eq:ConcMixed})), we directly arrive at the expression~(\ref{eq:ConcMixed}) for the lower bound~(\ref{app:ConcLowerBound}) in terms of randomized population probabilities
\begin{align}
&C_N(\varrho)^2 \nonumber  \geq \\
& \geq2^{2-N}-2^{2(1-N)}\sum_{A\subseteq \{1,\ldots,N\}}\text{tr}[\varrho_A^2]+(4-2^{2-N}) \tr[\varrho^2] \nonumber \\
&=2^{2-N}-2^{2(1-N)}d^N(d+1)^N\sum_{\vec s \in \{0,1\}^N}\mathbb E_U\left[{P_{U}(\vec{s})^2}\right] + \nonumber \\
&+(4-2^{2-N}) d^{N} \sum_{\mathbf{s},\mathbf{s}'} (-d)^{D(\mathbf{s}, \mathbf{s}')} \mathbb E_U\left[P_U(\mathbf{s})P_U(\mathbf{s}')\right] .
\label{eq:ConcMixedApp}
\end{align}

\subsection{Multiparticle concurrence as a function of the moments~(\ref{eq:RandomMomentsQudits})}\label{app:A2}
We further note that Eqs.~(\ref{eq:concpopprob}) and (\ref{eq:ConcMixedApp}) can alternatively be expressed as a function of the moments~(\ref{eq:RandomMomentsQudits}), introduced in Sec.~\ref{sec:RandMoments}. To do so, we remind the reader of the representation of $N$-particle quantum states in terms of its sector lengths, along the lines of Refs.~\cite{NikolaiSectorLengths,SatoyaMoments}. First, we note that a state $\varrho$ can always be  expressed as follows
\begin{align}
\varrho=\frac{1}{d^N}\sum_{i_1,\ldots,i_N=0}^{d^2-1} c_{i_1,\ldots,i_N} \lambda_{i_1}\otimes\ldots \otimes\lambda_{i_N}
\end{align}
where $\lambda_{0}$ is the identity, and 
$\lambda_{i}$
are the Gell-Mann matrices, normalized such that
$\lambda_i=\lambda_i^\dagger$,
$\mathrm{tr}\left[\lambda_{i} \lambda_{j}\right]=d \delta_{i j}$,
and $\mathrm{tr}\left[\lambda_{i} \right]=0$ for $i > 0$.
The real coefficients $c_{i_{1} \cdots i_{N}}$ are given by
$c_{i_{1} \cdots i_{N}}=\mathrm{tr} \left[\varrho \lambda_{i_{1}} \otimes \cdots \otimes \lambda_{i_{N}} \right]=\expec{\lambda_{i_{1}} \otimes \cdots \otimes \lambda_{i_{N}}}$.
The state $\varrho$ can be represented by
\begin{align}
    \varrho = \frac{1}{d^{N}} \left(
    \mathbb 1^{\otimes n} + \hat A_1 + \hat A_2 + \cdots + \hat A_N
    \right), 
\end{align}
where the Hermitian operators $\hat A_k$,
with $k=1,\ldots,N$, denote the sum of all terms
coming from the basis elements with weight $k$
\begin{align}
    \hat A_k(\varrho)= \sum_{\substack{i_{1}, \cdots, i_{N}=0,\\ \mathrm{wt}(\lambda_{i_{1}} \otimes \cdots \otimes \lambda_{i_{N}})=k }}^{d^2-1}
    c_{i_{1} \cdots i_{N}}
\lambda_{i_{1}} \otimes \cdots \otimes \lambda_{i_{N}},
\end{align}
where the weight $\mathrm{wt}(\lambda_{i_{1}} \otimes \cdots \otimes \lambda_{i_{N}})$
is equal to the number of non-identity Gell-Mann matrices in the product $\lambda_{i_{1}} \otimes \cdots \otimes \lambda_{i_{N}}$.
Now, we can define sector lengths as
\begin{align}  \label{eq-sectorlengths}
    A_k(\varrho) =\frac{1}{d^n}\mathrm{tr}\left[\hat A_k(\varrho)^2
    \right]
    = \sum_{\substack{i_{1}, \cdots, i_{N}=0,\\ \mathrm{wt}(\lambda_{i_{1}} \otimes \cdots \otimes \lambda_{i_{N}})=k }}^{d^2-1}
    c_{i_{1} \cdots i_{N}}^2. 
\end{align}
Physically, the sector lengths $A_k$ quantify the amount
of $k$-body quantum correlations. Note that $A_0=\alpha_{0 \cdots 0}=1$
due to $\text{tr}(\varrho)=1$. The sector lengths $A_k$ can be associated
with the purity
of $\varrho$:
\begin{align} \label{eq-sector-purity}
    \mathrm{tr}(\varrho^2) = \frac{1}{d^{N}} \sum_{i_{1}, \cdots, i_{N}=0}^{d^2-1} c_{i_{1} \cdots i_{N}}^2 = \frac{1}{d^{N}} \sum_{k=0}^{N}A_k(\varrho). 
\end{align}
Further on, as shown in Refs.~\cite{tran1,SatoyaMoments}, the sector lengths~(\ref{eq-sectorlengths}) are directly related to the moments~(\ref{eq:RandomMomentsQudits}) evaluated on the respective qudit sector:
\begin{align}
\frac{1}{(d^2-1)^k} A_k(\varrho)=\sum_{|A|=k}\mathcal R_A^{(2)}(\varrho),
\end{align} 
where $A=\{i_1,\ldots,i_k\}\subset \{1,\ldots,N\}$ with cardinality $k$. Hence, using Eq.~(\ref{eq-sector-purity}), the purity can be expressed as a sum of the second order moments evaluated on all possible subsets $A$, as follows
\begin{align} \label{eq-2nd-moments}
    \mathrm{tr}(\varrho^2)& = \frac{1}{d^{N}} \sum_{k=0}^{N}(d^2-1)^k\sum_{|A|=k}\mathcal R_A^{(2)}(\varrho)\nonumber \\ 
    &= \frac{1}{d^{N}} \sum_{A\subset \{1,\ldots,N\}}(d^2-1)^{|A|} \mathcal R_A^{(2)}(\varrho). 
\end{align}
Lastly, it is straightforward to express the pure-state concurrence~(\ref{eq:concurrence}) as a function of second order moments by invoking Eq.~(\ref{eq-2nd-moments}):
\begin{align}
C_N(\ket\psi) &= 2\sqrt{1-\sum_{A \subset \{1, \ldots, N \}}  \sum_{A' \subset A} \frac{(d^2-1)^{|A'|}}{2^Nd^{|A|}} \mathcal R^{(2)}_{A'}}\nonumber \\
= 2&\sqrt{(1-\frac{1}{2^N})-\sum_{A \subsetneq \{1, \ldots, N \}}  \sum_{A' \subset A} \frac{(d^2-1)^{|A'|}}{2^Nd^{|A|}} \mathcal R^{(2)}_{A'}}.
\label{app:concurrence_moments2}
\end{align}
Note that in Eq.~(\ref{app:concurrence_moments2}) we used that for pure states the purity of the total state is one which eliminates several summands, in particular, also those which are of cardinality $|A|=N$.

Evaluating Eq.~(\ref{app:concurrence_moments2}) for the special cases $N=2,3$ and $d=2$ leads to
\begin{align}
C_2(\ket\psi) &= \sqrt{1-\frac{3}{2}(\mathcal R^{(2)}_1+\mathcal R^{(2)}_2)} \nonumber  \\
C_3(\ket\psi) &= \frac{1}{\sqrt{2}} \left\{\frac{15}{4}-3(\mathcal R^{(2)}_1+\mathcal R^{(2)}_2+\mathcal R^{(2)}_3)\right. \nonumber \\
&\left.-\frac{3^2}{4}(\mathcal R^{(2)}_{12}+\mathcal R^{(2)}_{23}+\mathcal R^{(2)}_{13})\right\}^{\frac{1}{2}}.
\label{app:concurrence_moments_q3}
\end{align}
Furthermore, for mixed states, we can analogously use Eq.~(\ref{eq-2nd-moments}) to derive an expression of the lower bound~(\ref{eq:ConcLowerBound}) in terms of second moments only, yielding 
\begin{align}
&C_N(\varrho)^2= 2-2^{2-N} \times \nonumber \\
&\times \sum_{A \subset \{1, \ldots, N \}} \Big\{  \frac{1}{2^{|A|}} \sum_{A' \subset A} 3^{|A'|} \mathcal R^{(2)}_{A'} +\frac{2}{2^N} 3^{|A|} \mathcal R^{(2)}_A \Big\}.
\label{app:concurrence_bound_moments}
\end{align}

\section{Evaluation of statistical uncertainties}\label{app:UnbEst}

\subsection{Unbiased estimators and their variance}\label{app:UnbEst}
In an experiment one can estimate the population probabilities $P_U(\vec{s})$ (In the following we will sometimes drop the subscript $U$ unless it is required by the context.)  only from a finite number $K$ of projective measurements. Then, the corresponding statistical esimator is given by $\widetilde{P}(\boldsymbol s) = Y(\boldsymbol s)/K$, where $Y(\boldsymbol s)$ is the absolute frequency with which the bitstring $\vec{s}$ appears. Hence, the random variable $Y(\boldsymbol s)$ is distributed according to a multinomial distribution with probabilities $\{P(\vec{s})\}_{\vec s \in \{0,1\}^N}$ and $K$ trials. Exploiting this fact one can find unbiased estimators $\widetilde{P}^{(k)}(\boldsymbol s)$ for the $k$-th power of the population probability $P(\vec{s})^k$ by making the ansatz $\widetilde{P}^{(k)}(\boldsymbol s)=\sum_{i=0}^k \alpha_i {(\tilde Y(\boldsymbol s)/K)}^i$ with the condition that $\mathbb E_{\text{multi}}[{\widetilde{P}^{(k)}(\boldsymbol s)}]=P(\boldsymbol s)^k$. For the three lowest orders this results in

\begin{widetext}
\begin{align}
\widetilde{P}^{(2)}(\boldsymbol s)&=\frac{\widetilde{P}(\boldsymbol s)(K\widetilde{P}(\boldsymbol s)-1)}{K-1}=\widetilde{P}(\boldsymbol s)\times \frac{K\widetilde{P}(\boldsymbol s)-1}{K-1},\\
\widetilde{P}^{(3)}(\boldsymbol s)&=\frac{ \widetilde{P}(\boldsymbol s)(K \widetilde{P}(\boldsymbol s)-1)(K \widetilde{P}(\boldsymbol s)-2)}{(K-1)(K-2)}=\widetilde{P}^{(2)}(\boldsymbol s)\times\frac{K \widetilde{P}(\boldsymbol s)-2}{K-2},\\
\widetilde{P}^{(4)}(\boldsymbol s)&=\frac{\widetilde{P}(\boldsymbol s)(K \widetilde{P}(\boldsymbol s)-1)(K \widetilde{P}(\boldsymbol s)-2)(K \widetilde{P}(\boldsymbol s)-3)}{(K-1)(K-2)(K-3)} =\widetilde{P}^{(3)}(\boldsymbol s)\times\frac{K \widetilde{P}(\boldsymbol s)-3}{K-3},
\end{align}
\end{widetext}

and similarly for products of population probabilities $P(\vec s)P(\vec{s}')$, with $\vec{s}\neq \vec{s}'$, we obtain 
\begin{align}
\widetilde{P}^{(1,1)}(\vec{s},\vec{s}') = \frac{Y(\vec{s})  Y(\vec{s}')}{K(K-1)}=\frac{K}{K-1}\widetilde{P}(\boldsymbol s)\widetilde{P}(\boldsymbol s').
\label{eq:est_product}
\end{align}

Moreover, also the expectation value $\expi{\ldots}$ taken with respect to the local measurement settings can only be estimated based on finite samples of measurement bases which finally leads to the definition of the unbiased estimators reported in Eqs.~(\ref{eq:EstPs2}) and (\ref{eq:EstPsPsp}) of the main text. 

Further on, we have to investigate the variance of the estimators~(\ref{eq:EstPs2}) and (\ref{eq:EstPsPsp}) in order to get a handle on their associated statistical error. We will start with the calculation of the variance of  Eq.~(\ref{eq:EstPs2}) which reads
\begin{align}
\Var{\overline{P^2_U(\vec{s})}} &= \frac{1}{M^2} \sum_{i=1}^{M} \Var{\widetilde{P}_{U_i}^{(2)}(\boldsymbol s)}  \nonumber \\
&= \frac{1}{M} \expubi{\big(\widetilde{P}_{U_i}^{(2)}(\boldsymbol s)\big)^2} - \expi{P(\vec s)^2}^2,
\label{app:VarEstPs2_1}
\end{align}
where we used that individual samples of local unitary transformations $U_i$ are independent and identically distributed (i.i.d.).  In order to further evaluate the expression~(\ref{app:VarEstPs2_1}) we have to exploit the following moments of the multinomial distribution:

\begin{widetext}
\begin{align}
\expbi{\widetilde{P}^{(2)}(\boldsymbol s)} &= \onefrac{K^2} \, \expbi{Y(\boldsymbol s)^2} = \onefrac{K} \left[(K-1)P(\vec s)^2 - P(\vec s)\right],\\
\expbi{\widetilde{P}^{(3)}(\boldsymbol s)} &= \onefrac{K^2} \begin{aligned}[t] &\big[(K-1)(K-2)P(\vec s)^3   +3(K-1)P(\vec s)^2 + P(\vec s)\big],  \end{aligned}\\
\expbi{\widetilde{P}^{(4)}(\boldsymbol s)} &= \onefrac{K^3} \begin{aligned}[t]
\big[&(K-1)(K-2)(K-3)P(\vec s)^4  + 6(K-1)(K-2)P(\vec s) ^3 + 7(K-1)P(\vec s)^2 + P(\vec s)\big],\end{aligned}
\end{align}

leading to

\begin{align}
\expubi{\big(\widetilde{P}^{(2)}(\boldsymbol s)\big)^2} &= \expubi{\frac{K^2 \widetilde{P}(\boldsymbol s)^4 -2K\widetilde{P}(\boldsymbol s)^3 + \widetilde{P}(\boldsymbol s)^2}{(K-1)^2}}\\
&= \frac{\expi{K^2 \, \expbi{\widetilde{P}(\boldsymbol s)^4} -2K \, \expbi{\widetilde{P}(\boldsymbol s)^3} + \expbi{\widetilde{P}(\boldsymbol s)^2}}}{(K-1)^2}\nonumber\\
&= \frac{\expi{(K-2)(K-3)P(\vec s)^4 + 4(K-2)P(\vec s)^3 + 2P(\vec s)^2}}{K(K-1)}. 
\end{align}
\end{widetext}
%
%
Now we can write the variance as follows
\begin{align}\label{app:VarP2}
\Var{\overline{P^2_U(\vec{s})}} = &\frac{1}{{MK(K-1)}} \Big[(K-2)(K-3) \, \expi{P^4_U(\vec{s})} \nonumber\\
&+ 4(K-2) \, \expi{P^3_U(\vec{s})} + 2 \, \expi{P^2_U(\vec{s})} \Big] \nonumber\\
&- \frac{\expi{P^2_U(\vec{s})}^2}{M}.
\end{align} 
An analogous calculation can be applied in case of the estimator~(\ref{eq:EstPsPsp}) leading to the following expression for its variance 
\begin{align}
&\Var{\overline{P_U(\vec{s})P_U(\vec{s}')}} = \frac{1}{MK(K-1)}\times \nonumber\\
&\times \bigg\{
(K-2)(K-3) \expi{P_U^2(\vec{s})P_U^2(\vec{s}')} + \nonumber \\
& (K-2) \left(\expi{P_U^2(\vec{s})P_U(\vec{s}')} + \expi{P_U(\vec{s})P_U^2(\vec{s}')}\right) \nonumber \\
&+ \expi{P_U(\vec{s})P_U(\vec{s}')}\bigg\} - \frac{1}{M} \expi{{P_{U}(\vec{s})P_{U}(\vec{s}')}}^2.
\label{app:VarPsPsp}
\end{align}
We note that the all the (cross-)moments contained in Eqs.~(\ref{app:VarP2}) and (\ref{app:VarPsPsp}) can be evaluated using the following expression
\begin{align}
\expi{P^t_U(\vec{s})P^k_U(\vec{s}')} &= D_{d,t}^N \tr{\Bigg[\varrho^{\otimes t} \bigotimes_{i=1}^N\left.\begin{cases}
	P_+^{(i)}, & s_i = s'_i\\
	{A}_{t,k}^{(i)}, & s_i \neq s'_i 
	\end{cases}\right\}}\Bigg],
\label{app:AvPsPspU}
\end{align}
where $P_+^{(i)}$ denotes the projector onto the symmetric subspace (see Eq.~(\ref{app:ProjectorSymt})) acting on the $i$-th particle and ${A}_{t,k}^{(i)}$ is defined as 
\begin{equation}
{A}_{t,k}^{(i)} = \expui{u_i^{\otimes (t+k)} \ketbra{0}{0}^{\otimes t}\otimes \ketbra{1}{1}^{\otimes k} \left(u_i^\dagger\right)^{\otimes (t+k)}}.
\end{equation} 
Note that from Eq.~\ref{app:AvPsPspU} it becomes clear that the variance~(\ref{app:VarPsPsp}) depends on the Hamming distance between the bitstrings $\vec{s}$ and $\vec{s}'$.

In conclusion, we see that the information about the statistical error of the estimators~(\ref{eq:EstPs2}) and (\ref{eq:EstPsPsp}) is encoded in the (cross-)moments~(\ref{app:AvPsPspU}). In the following we will evaluate these quantities exactly for a set of typical multiparticle states, i.e., the $N$ qubit ground state, the GHZ state and for Haar random states, and derive symbolic expressions that hold for an arbitrary number of particles $N$. 

\subsection{Variance of typical multiparticle states}
\subsubsection{Product State}
For a random $N$-qubit product state we have $\varrho^{\otimes t} = \bigotimes_{i=1}^N \ketbra{\varphi_i}{\varphi_i}^{\otimes t}$ leading directly to the  expressions:
\begin{align}
\expi{P^t_U(\vec{s})} &= D_{d,t}^N \trc{\bigotimes_{i=1}^N \ketbra{\varphi_i}{\varphi_i}^{\otimes t} \, \bigotimes_{i=1}^N P_+^{(i)}}\nonumber \\
&= D_{d,t}^N \prod_{i=1}^N \trc{\ketbra{\varphi_i}{\varphi_i}^{\otimes t} P_+^{(i)}}\nonumber \\
&= D_{d,t}^N  \trc{\ketbra{0}{0}^{\otimes t} P_+}^N,
\end{align}
which can be evaluated easily using Eq.~(\ref{eq:HaarIntU}).

\subsubsection{GHZ State}
For the $N$-qubit GHZ state we have
\begin{align}
\varrho = \onefrac{2} \sum_{\vec{i}, \vec{i}'} \, (\delta_{\vec{i}, \vec{0}}\delta_{\vec{i'}, \vec{0}} + \delta_{\vec{i}, \vec{0}}\delta_{\vec{i'}, \vec{1}} + \delta_{\vec{i}, \vec{1}}\delta_{\vec{i'}, \vec{0}} + \delta_{\vec{i}, \vec{1}}\delta_{\vec{i'}, \vec{1}}) \, \ketbra{\vec{i}}{\vec{i}'},
\end{align}
which can be directly used to analytically evaluate  the expectation value of the higher order population probabilities 
\begin{align}
\expi{P^t_U(\vec{s})} &= \frac{ D_{d,t}^N }{2^t} \left[ \left([P_+]_{0\dots 0, 0\dots 0}\right)^N + \left([P_+]_{0\dots 1, 0\dots 0}\right)^N \right.\nonumber \\
&\left.+ \dots + \left([P_+]_{1\dots 1, 1\dots 1}\right)^N\right].
\end{align}
where $[P_+]_{\vec i, \vec{i}'}$ denotes the respective matrix element $(\vec i,\vec{i}')$ of the projector $P_+$. 
Analogously, we find for the cross-terms
\begin{align}
&\expi{P^t_U(\vec{s})P^k_U(\vec{s}')} \nonumber \\
&= \frac{ D_{d,t}^{N}}{2^t} \bigg[ \left([P_+]_{0\dots 0, 0\dots 0}\right)^{N-D(\mathbf{s}, \mathbf{s}')}\left({[A^{t,k}]}_{0\dots 0, 0\dots 0}\right)^{D(\mathbf{s}, \mathbf{s}')} \nonumber\\
&+ \left([P_+]_{0\dots 1, 0\dots 0}\right)^{N-D(\mathbf{s}, \mathbf{s}')}\left({[A^{t,k}]}_{0\dots 1, 0\dots 0}\right)^ {D(\mathbf{s}, \mathbf{s}')} + \dots  \nonumber\\
&+ \left([P_+]_{1\dots 1, 1\dots 1}\right)^{N-D(\mathbf{s}, \mathbf{s}')}\left({[A^{t,k}]}_{1\dots 1, 1\dots 1}\right)^{D(\mathbf{s}, \mathbf{s}')}\bigg].
\end{align}

\subsubsection{Haar Random States}
Finally, for Haar random states we evaluate the expected average variance as 
\begin{align}
&\exppsi{\Var{\overline{P^2_U(\vec{s})}}} = \, \frac{1}{MK(K-1)} \times \nonumber \\
&\times \bigg\{ (K-2)(K-3) \, \exppsiu{P^4_U(\vec{s})} + 4(K-2) \exppsiu{P^3_U(\vec{s})} \nonumber\\
&+ 2 \, \exppsiu{P^2_U(\vec{s})}\bigg\} - \frac{1}{M} \, \exppsi{\expi{P^2_U(\vec{s})}^2}.
\end{align}
which can be evaluated straightforwardly using Eq.~(\ref{eq:HaarIntU}).

As $\expi{P^2_U(\vec{s})}$ does not depend on the specific bitstring under consideration, to improve the statistics in a real experiment  one can also average over all different bit-strings that were observed, i.e., $I=\{s_1, \dots, s_{|I|}\}\subset \{0,1\}^N$ by considering the estimator introduced in Eq.~(\ref{eq:EstCinclAvOfBitstrings}) of the main text.  Here, in general $|I| \leq K$  as maximally $K$ different bitstrings can be measured during $K$ projective measurements. As for $N$ qubits the maximal number of different possible bitstrings is given by $2^N$ we will estimate the number of different bitstrings that occured by $|I| = \min\left(2^N, \, K\right)$ which is a rough estimate justified for the specific choice of Haar random states. Having this in mind we consider the variance of the estimator~(\ref{eq:EstCinclAvOfBitstrings}), yielding
\begin{align}
\Var{\overline{P^2_U}} &= \frac{1}{M^2} \sum_{i=1}^{M} \frac{1}{|I|^2} \sum_{\vec{s}, \vec{s}'} \text{Cov}\left[\tilde P^2_U(\vec{s}), \tilde P^2_U(\vec{s}')\right]\nonumber\\
&= \frac{1}{M^2} \sum_{i=1}^{M} \frac{1}{|I|^2} \left( \sum_{\vec{s}} \Var{\left(\tilde P^2_U(\vec{s})\right)_i}\right.\nonumber \\
 &\left.+ \sum_{\vec{s} \neq \vec{s}'} \text{Cov}\left[\left(\tilde P^2_U(\vec{s}), \tilde P^2_U(\vec{s}')\right)_i\right]\right).
\label{eq:haar_diffs}
\end{align}
In general, the term $\text{Cov}\left[\tilde P^2_U(\vec{s}), \tilde P^2_U(\vec{s}')\right]$ is difficult to evaluate analytically. However, for Haar random states the correlations between the probabilities of different outcomes $\vec{s}$ and $\vec{s}'$ are exponentially small in the number of qubits \cite{RandomMatrixStuff1,RandomMatrixStuff2}. Therefore, in the case of Haar random states and large system sizes it is well justified to approximate the variance as follows
\begin{align}
\Var{\overline{P^2_U}} \approx \frac{1}{M} \frac{1}{|I|} \Var{\overline{P^2_U(\vec{s})}}.
\end{align}
Similarly for the product term we can also determine the expected variance,
\begin{widetext}
\begin{align}
\exppsi{\Var{\overline{P_U(\vec{s})P_U(\vec{s}')}}} &= \, \frac{1}{MK(K-1)} \bigg\{(K-2)(K-3) \exppsiu{P_U^2(\vec{s})P_U^2(\vec{s}')} \nonumber  \\
&+ (K-2) \bigg(\exppsiu{P_U^2(\vec{s})P_U(\vec{s}')}+ \exppsiu{P_U(\vec{s})P_U^2(\vec{s}')}\bigg)\nonumber \\
&+ \exppsiu{P_U(\vec{s})P_U(\vec{s}')}\bigg\}- \frac{1}{M} \exppsi{\expi{{P_{U}(\vec{s})P_{U}(\vec{s}')}}^2}.
\end{align}
\end{widetext}
Here, $\exppsiu{P_U^n(\vec{s})P_U^m(\vec{s}')}$ can be evaluated once again using Eq.~(\ref{eq:HaarIntU}).  However, note that $\exppsi{\Var{\overline{P_U(\vec{s})P_U(\vec{s}')}}}$ depends explicitly on the specific bitstrings $\vec{s}$ and $\vec{s}'$.


\begin{thebibliography}{99}
\bibliographystyle{unsrt}

\bibitem{NielsenChuang} M. A. Nielsen and I. L. Chuang, \textit{Quantum Computation and Quantum Information} \href{https://doi.org/10.1017/CBO9780511976667}{(Cambridge University Press, New York, 2000)}.

\bibitem{Buhrman:2010uq} H. Buhrman, R. Cleve, S. Massar, R. de Wolf, Ronald, Rev. Mod. Phys., \textbf{82}, 665 (2010).


\bibitem{QuSup} F. Arute, et al., Nature \textbf{574}, 505 (2019).

\bibitem{SupercondQubitsRev} M. Kjaergaard, M. E. Schwartz, J. Braumüller, P. Krantz, Philip and J. I. -J.Wang,  S. Gustavsson, and W. D. Oliver, Annu. Rev. Condens. Matter Phys. \textbf{11}, 369 (2020).

\bibitem{Graselli2021} F. Graselli et al., arXiv:2111.05363.

\bibitem{Gessner2021} Z. Ren, W. Li, A. Smerzi, and M. Gessner, Phys.Rev.Lett. \textbf{126}, 080502 (2021).

\bibitem{VanDenNest2013} M. Van den Nest, Phys. Rev. Lett. \textbf{110}, 060504 (2013).

\bibitem{H4EntReview} R. Horodecki, P. Horodecki, M. Horodecki, K. Horodecki, Rev. Mod. Phys. \textbf{81}, 865 (2009).

\bibitem{OtfriedReview} O. G\"uhne, and G. T\'oth, Phys. Rep. \textbf{474}, 1 (2009).

\bibitem{ReviewCertification} J. Eisert, D. Hangleiter, N. Walk, I. Roth, D. Markham, R. Parekh, U. Chabaud, and E. Kashefi, Nat. Rev. Phys. {\bf 2}, 
382 (2020). 

\bibitem{TomographyReview} R. Blume-Kohout, New J. Phys. \textbf{12}, 043034 (2010).

\bibitem{SelfTestingRev} I. \u{S}upi\'c and J. Bowles, Quantum \textbf{4}, 337 (2020).

\bibitem{FidelityWitnesses} M. Gluza, M. Kliesch, J. Eisert, and L. Aolita, Phys. Rev. Lett. \textbf{120}, 190501 (2018). 

\bibitem{CompressedSensing1} D. Gross, Y.-K. Liu, S. T. Flammia, S. Becker, and J. Eisert, Phys. Rev. Lett. \textbf{105}, 150401 (2010).

\bibitem{CompressedSensing2} C. A. Riofrío, D. Gross, S. T. Flammia, T. Monz, D. Nigg, R. Blatt, J. Eisert, Nat. Comm. \textbf{8}, 15305 (2017).


\bibitem{LeandroRegenerativeModels} J. Carrasquilla, G. Torlai, R. G. Melko and L. Aolita, Nature Machine Intelligence \textbf{1}, 155 (2019).

\bibitem{RudolphRandMeasBell} Y.-C. Liang, N. Harrigan, S. D. Bartlett, and T. Rudolph, Phys. Rev. Lett. \textbf{104}, 050401 (2010).

\bibitem{FlammiaFidelityStat} S. T. Flammia and Y.-K. Liu, Phys. Rev. Lett. \textbf{106}, 230501 (2011).

\bibitem{BrunnerRandMeasBell} P. Shadbolt, T. V\'ertesi, Y.-C. Liang, C. Branciard, N. Brunner, and J. L. O'Brien, Sci. Rep. \textbf 2, 470 (2012).

\bibitem{EnkPRL2012} S. J. van Enk and C. W. J. Beenakker, Phys. Rev. Lett. \textbf{108}, 110503 (2012).

\bibitem{tran1} M. C. Tran, B. Daki\'c, F. Arnault, W. Laskowski, and T. Paterek, Phys. Rev. A \textbf{92}, 050301(R) (2015).

\bibitem{tran2} M. C. Tran, B. Daki\'c, W. Laskowski, and T. Paterek, Phys. Rev. A \textbf{94}, 042302 (2016).

\bibitem{Mattia2016} M. Walschaers, J. Kuipers, J.-D. Urbina, K. Mayer, M. C. Tichy, K. Richter and A. Buchleitner, New J. Phys. \textbf{18}, 032001 (2016).

\bibitem{Giordani2018} T. Giordani, et al., Nature Photonics \textbf{12}, 173 (2018).

\bibitem{MeMoments1} A. Ketterer, N. Wyderka, and O. G\"uhne, Phys. Rev. Lett. \textbf{122}, 120505 (2019).

\bibitem{MeMoments2} A. Ketterer, N. Wyderka, and O. G\"uhne, Quantum \textbf{4}, 325 (2020).

\bibitem{MichaelBachelor} M. Krebsbach, Bachelor thesis, Albert-Ludwigs-Universit\"at Freiburg (2019); https://freidok.uni-freiburg.de/data/150706.

\bibitem{ZollerFirst} A. Elben, B. Vermersch, M. Dalmonte, J. I. Cirac, and P. Zoller, Phys. Rev. Lett. \textbf{120}, 050406 (2018).

\bibitem{vermerschPRA} B. Vermersch, A. Elben, M. Dalmonte, J. I. Cirac, 
and P. Zoller, Phys. Rev. A {\bf 97}, 023604 (2018). 

\bibitem{ZollerScience} T. Brydges, A. Elben, P. Jurcevic, B. Vermersch, C. Maier, B. P. Lanyon, P. Zoller, R. Blatt, C. F. Roos, Science \textbf{364}, 260 (2019).

\bibitem{ElbenPRA} A. Elben, B. Vermersch, C. F. Roos, P.  Zoller, Phys. Rev. A \textbf{99}, 052323 (2019).

\bibitem{DakicRandomMeas} V. Saggio, A. Dimić, C. Greganti, L. A. Rozema, P. Walther, B. Daki\'c, Nat. Phys. \textbf{15}, 935 (2019).

\bibitem{MeineckeExperimentRandom} L. Knips, J. Dziewior, W. K\l obus, W. Laskowski, T. Paterek, P. J. Shadbolt, H. Weinfurter, and J. D. A. Meinecke, npj Quantum Information \textbf 6, 51 (2020).

\bibitem{ElbenPRL} A. Elben, B. Vermersch, R. van Bijnen, C. Kokail, T. Brydges, C. Maier, M. K. Joshi, R. Blatt, C. F. Roos, and P. Zoller, Phys. Rev. Lett. \textbf{124}, 010504 (2020).

\bibitem{ElbenPRLmixedstate} A. Elben, R. Kueng, H.-Y. Huang, R. van Bijnen, C. Kokail, M. Dalmonte, P. Calabrese, B. Kraus, J. Preskill, P. Zoller, B. Vermersch, Phys. Rev. Lett. \textbf{125}, 200501 (2020).

\bibitem{RandTriads} S.-X. Yang, G. N. Tabia, P.-S. Lin, and Y.-C. Liang, Phys. Rev. A \textbf{102}, 022419 (2020).

\bibitem{SatoyaMoments} S. Imai, N. Wyderka, A. Ketterer, and O. G\"uhne, Phys. Rev. Lett. \textbf{126}, 150501 (2021).

\bibitem{MeMoments3} A. Ketterer, S. Imai, N. Wyderka, and O. G\"uhne, Phys. Rev. A 106, L010402 (2022).

\bibitem{KnipsPerspective} L. Knips, Quantum Views \textbf 4, 47 (2020).


\bibitem{Shadows1} S. Aaronson, in Proceedings of the 50th Annual ACM SIGACT Symposium on Theory of Computing, (ACM, New York, 2018).

\bibitem{Shadows2} H.-Y. Huang, R. Kueng, and J. Preskill, Nature Physics \textbf{16}, 1050 (2020).

\bibitem{RandomMeasTomography} M. Paini, A. Kalev, D. Padilha, and B. Ruck, arXiv:2011.04754v2.

\bibitem{NikolaiSectorLengths} N. Wyderka and O. G\"uhne, J. Phys. A: Math. Theor. \textbf{53}, 345302 (2020).

\bibitem{Mintert2004} F. Mintert, M. Ku\'s, and A. Buchleitner, Phys. Rev. Lett. \textbf{92}, 167902 (2004).


\bibitem{Carvalho2004} A. R. R. Carvalho, F. Mintert, and A. Buchleitner, Phys. Rev. Lett. \textbf{93}, 230501 (2004).

\bibitem{Mintert2005a} F. Mintert, A. R. R. Carvalho, M. Ku\'s, and A. Buchleitner, Physics Reports \textbf{415}, 207 (2005).

\bibitem{MileGu2021RandMeas} Z. Liu, P. Zeng, Y. Zhou, and M. Gu, Phys. Rev. A: \textbf{105}, 022407 (2022). 

\bibitem{MScSophia} S.~Ohnemus, Master~Thesis, Albert-Ludwigs-Universit\"at~Freiburg~(2021); https://doi.org/10.6094/UNIFR/227071


\bibitem{Mintert2005b} F. Mintert, M. Ku\'s, and A. Buchleitner, Phys. Rev. Lett. \textbf{93}, 230501 (2005).

\bibitem{AolitaConc1} Leandro Aolita and Florian Mintert, Phys. Rev. Lett. \textbf{97}, 050501 (2006).

\bibitem{AolitaConc2} Leandro Aolita, Andreas Buchleitner and Florian Mintert, Phys. Rev. A \textbf{78}, 022308 (2008).

\bibitem{Dankert} C. Dankert, M.Sc. thesis, University of Waterloo, (2005); also available as e-print \href{https://arxiv.org/abs/quant-ph/0512217}{quant-ph/0512217}.

\bibitem{existence} P. D. Seymour, T. Zaslavsky, \href{https://doi.org/10.1016/0001-8708(84)90022-7
}{Advances in Mathematics \textbf{52}, 213 (1984)}.

\bibitem{ApproxDesign1} F. G. S. L. Brand\~ao, A. W. Harrow, and M. Horodecki, \href{http://doi.org/10.1103/PhysRevLett.116.170502
}{Phys. Rev. Lett. \textbf{116}, 170502 (2016)}.

\bibitem{ApproxDesign2} Y. Nakata, C. Hirche, M. Koashi, A. Winter, \href{https://doi.org/10.1103/PhysRevX.7.021006
}{Phys. Rev. X \textbf 7, 021006 (2017)}.

\bibitem{ApproxDesign3} J. Haferkamp, F. Montealegre-Mora, M. Heinrich, J. Eisert, D. Gross, and I. Roth, \href{https://arxiv.org/abs/2002.09524}{arXiv:2002.09524}.

\bibitem{Cliff3Design} Z. Webb, \href{https://doi.org/10.26421/QIC16.15-16}{Quantum Inf. Comput. \textbf{16}, 1379 (2016)}.

\bibitem{CliffNo4Design} H. Zhu, R. Kueng, M. Grassl, and D. Gross, \href{https://arxiv.org/abs/1609.08172}{arXiv:1609.08172}.



\bibitem{Ullah1946} N. Ullah, Nuclear Physics \textbf{58}, 65 (1964).

\bibitem{Petz2004} D. Petz and J. Rffy, Period. Math. Hungar. \textbf{49}, 103 (2004).

\bibitem{SchmidtSpringer2010} K. D. Schmidt, \textit{Ma\ss \ und Wahrscheinlichkeit}, \href{https://doi.org/10.1007/978-3-642-21026-6}{(Springer-Verlag, Heidelberg, 2011)}.

\bibitem{Chamon1} D. Shaffer, C. Chamon, A. Hamma and E. R Mucciolo, Phys. Rev. Lett. \textbf{112}, 240501 (2014). 

\bibitem{Chamon2} D. Shaffer, C. Chamon, A. Hamma and E. R Mucciolo, J. Stat. Mech.  P12007 (2014). 

\bibitem{NandoRandomCirc} Raul O. Vallejos, Fernando de Melo, and Gabriel G. Carlo, Phys. Rev. A \textbf{104}, 012602 (2021).

\bibitem{GottesmanKnillTheo} D. Gottesman, The Heisenberg representation of quantum computers, Group22: Proceedings of the XXII International Colloquium on Group Theoretical Methods in Physics, edited by S. P. Corney, R. Delbourgo, and P. D. Jarvis (International Press, Cambridge, MA, 1999), pp. 32–43.

\bibitem{CliffClark2008} S. Clark, R. Jozsa, and N. Linden, 
Quantum Inf. Comput. \textbf{8}, 106 (2008).

\bibitem{CliffJosza2014} R. Jozsa and M. Van den Nest, Quantum Inf. Comput. \textbf{14}, 633 (2014).

\bibitem{MG_0} L. Valiant, 
SIAM J. Comput. \textbf{31}, 1229 (2002).

\bibitem{MG_1} R. Jozsa and A. Miyake, Proc. R. Soc. A \textbf{464}, 3089 (2008).

\bibitem{MG_2} D. J. Brod and A. M. Childs, Quantum Info. Comput. \textbf{14}, 901 (2014).

\bibitem{TerhalDivincenzo2002} B. M. Terhal and D. P. DiVincenzo, Phys. Rev. A \textbf{65}, 032325 (2002). 

\bibitem{IQP1} D. Shepherd and M. J. Bremner, Proc. R. Soc. A \textbf{465}, 1413 (2009).

\bibitem{IQP2} M. J. Bremner, R. Josza, and D. Shepherd, Proc. R. Soc. A \textbf{467}, 459 (2010).

\bibitem{PhaseRandomCircIQP} Y. Nakata and M. Murao, 
Eur. Phys. J. Plus \textbf{129}, 152 (2014).

\bibitem{IQP3} K. Fujii and T. Morimae, 
New J. Phys. \textbf{19}, 033003 (2017).


\bibitem{HaarAverages}  B. Collins and P. Sniady, Commun. Math. Phys. \textbf{264}, 773 (2006).

\bibitem{Weingarten} D. Weingarten, Journal of Mathematical Physics \textbf{19}, 999 (1978).

\bibitem{HaarConcentration1} M. Ledoux, \textit{The Concentration of Measure Phenomenon}, (Providence, RI: American Mathematical Society) (2001).

\bibitem{HaarConcentration2} M. Tiersch, et al., J. Phys. A: Math. Theor. \textbf{46}, 085301 (2013).





\bibitem{RandomMatrixStuff1} N. Ullah, Nuclear Physics \textbf{58}, 65 (1964).

\bibitem{RandomMatrixStuff2} D. Petz and J. Rffy, Period. Math. Hungar. \textbf{49}, 103 (2004).

\bibitem{CliffNo4Design} H. Zhu, R. Kueng, M. Grassl, and D. Gross, arXiv:1609.08172.





\end{thebibliography}
\end{document}